\documentclass{article}

 \usepackage[main, final]{neurips_2026}

\usepackage[utf8]{inputenc}
\usepackage[T1]{fontenc}

\usepackage{amsmath,amsfonts,amssymb}
\usepackage{graphicx}
\usepackage{booktabs}
\usepackage{multirow}
\usepackage{makecell}
\usepackage{tcolorbox}
\usepackage{array}
\usepackage{tabularx}
\usepackage{threeparttable}
\usepackage{colortbl}
\usepackage{xcolor}
\usepackage{subcaption}
\usepackage{caption}
\usepackage{adjustbox}
\usepackage{enumitem}
\usepackage{pifont}
\usepackage{ulem}
\usepackage{url}
\usepackage{microtype}
\usepackage{nicefrac}
\usepackage{siunitx}
\usepackage{graphicx}

\usepackage[ruled,vlined]{algorithm2e}

\usepackage{hyperref}
\hypersetup{hidelinks}

\newcommand{\best}[1]{\cellcolor{orange!15}\textbf{#1}}

\title{\textit{Speech Tokenizers are Vulnerable}: \\ Transferable Semantic Attack and Robust Tokenizer}

\author{%
  Zhisheng Zhang$^{1\dag}$, Yifan Mi$^{2\dag}$, Yixuan Zhou$^{1}$, Haiyun Li$^{1}$, Shengbo Cai$^{1}$, \\
  \textbf{Dongfu Song}$^{1}$, \textbf{Ziang Xu}$^{1}$, \textbf{Jie Hao}$^{2\ddag}$, \textbf{Zhiyong Wu}$^{1\ddag}$ \\
  $^1$ Shenzhen International Graduate School, Tsinghua University\\
  $^2$ Beijing University of Posts and Telecommunications \\ 
  \texttt{zhangzs25@mails.tsinghua.edu.cn}
}

\begin{document}
\maketitle

\renewcommand{\thefootnote}{\fnsymbol{footnote}}
\footnotetext[2]{Equal contribution.}
\footnotetext[3]{Corresponding authors.}
\renewcommand{\thefootnote}{\arabic{footnote}}

\begin{abstract}
  Speech tokenizers serve as speech front ends, encoding input speech signals into representations for downstream models, such as automatic speech recognition (ASR) systems and large audio-language models (LALMs).
  In this paper, we find that speech tokenizers are fragile at the semantic level: small perturbations can disrupt their semantic encoding process and consequently cause severe degradation in downstream ASR and LALM performance. To study this vulnerability, we propose \textbf{T-SemAttack}, a transferable semantic attack based on a semantic encoder ensemble. By jointly drifting the representation spaces of multiple semantic encoders, T-SemAttack effectively disrupts the semantic content of speech while preserving perceptual quality, thereby inducing error features.
  We further analyze layer-wise representation drift and cross-model gradient alignment. Our findings show that perturbations can induce substantial shifts in deep semantic representations and that the input gradients of different surrogate encoders exhibit directional alignment. These observations suggest that shared adversarially sensitive directions in the input space may be a key factor underlying cross-model attack transferability.
  Building on these, we introduce \textbf{ROSETok}, a robust speech tokenizer that combines noisy training with robust semantic distillation to improve reconstruction fidelity and downstream SpeechLLM robustness.
  Extensive experiments on multiple datasets, 9 open-source and black-box ASR systems, and 4 LALMs demonstrate that T-SemAttack achieves strong transferable attack performance, while the proposed Robust Speech Tokenizer exhibits robustness under high-fidelity reconstruction and downstream tasks. Our code and demo are available at \href{https://wxzyd123.github.io/t-semattack}{https://wxzyd123.github.io/t-semattack}.
\end{abstract}

\section{Introduction}
In recent years, LALMs~\cite{qwen2-audio, qwen3-omni} have demonstrated exceptional interaction capabilities and are widely deployed in scenarios, e.g., in-vehicle assistants and meeting transcription. Compared to previous speech models~\cite{vits, tacotron}, current large language model (LLM)-based architectures~\cite{qwen2-audio, af3, qwen3-asr} typically rely on a speech tokenizer to map continuous signals into discrete tokens, which are subsequently fed into LLMs for modeling. Consequently, the speech tokenizer bridges the continuous and discrete representations, and its robustness directly dictates the reliability of the entire semantic pipeline.

\begin{figure*}[t]
      \centering
      \includegraphics[width=\textwidth]{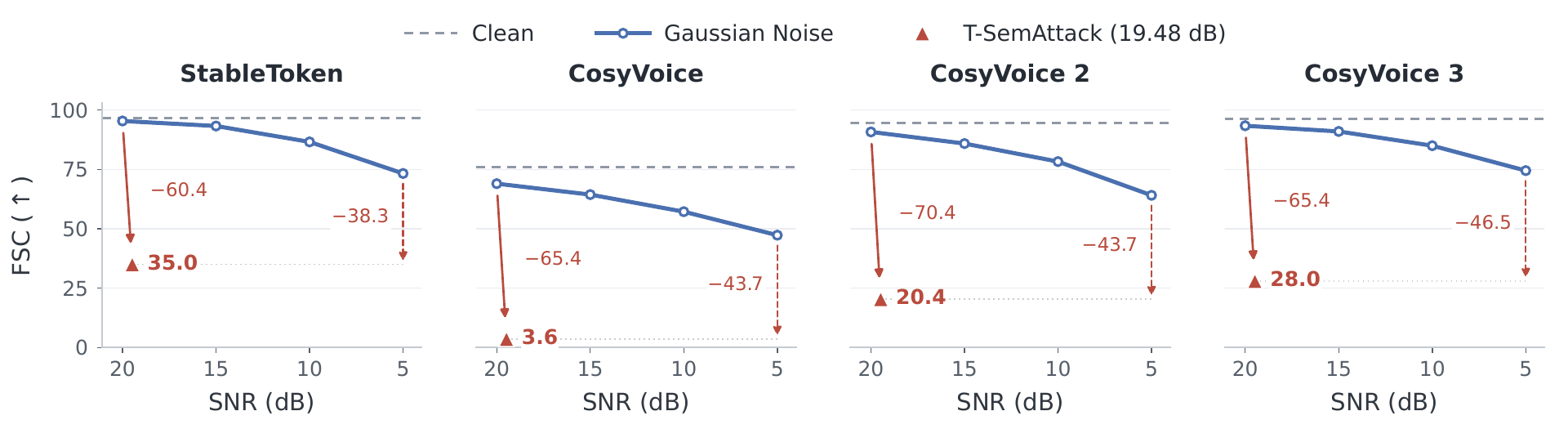}
      \caption{Gaussian Noise vs. T-SemAttack across Tokenizers. We feed the pre-quantized representations from each tokenizer into \textbf{SmolLM2-135M}~\cite{smollm2} and evaluate downstream \textbf{semantic understanding on the FSC spoken command intent recognition task (↑)}.}
      \label{fig:attack_comparison}
\end{figure*}

Existing studies primarily assess the vulnerability of speech tokenizers to Gaussian noise~\cite{stabletoken}, whereas semantic attacks can expose more severe vulnerabilities. As shown in Figure \ref{fig:attack_comparison}, the semantic attack (T-SemAttack) causes substantially greater degradation in downstream FSC performance across models than Gaussian noise, yielding scores even lower than those obtained under stronger Gaussian noise at an SNR of 5 dB. These results indicate that tolerance to random noise does not guarantee robustness against perturbations optimized to disrupt semantic representations.

In this paper, we propose a \textbf{T}ransferable \textbf{Sem}antic \textbf{Attack} (\textbf{T-SemAttack}), which focuses representation drifting in the speech tokenizer. By jointly perturbing multiple surrogate semantic encoders~\cite{zq-attack}, the method directly corrupts the semantic content of the input speech. We employ the psychoacoustic model~\cite{e2e-vguard} to preserve the perturbative perceptual quality. The resulting adversarial examples can therefore effectively interfere with the encoding process of the speech tokenizer while maintaining high perceptual quality, leading to severe semantic deviations in both ASR and LALM systems.

Furthermore, we analyze the effectiveness and transfer mechanisms of T-SemAttack from three perspectives: layer-wise representation drift, shared attack directions across models, and perturbation magnitude. We investigate how perturbations affect deep representations, whether shared sensitive directions underpin cross-model transferability, and how representation shifts translate into token changes and downstream recognition errors as perturbation magnitude increases. These analyses inform the training of tokenizers for semantic stability.

We consider the speech tokenizer as the safety bottleneck in ASR or LALMs. Based on our analysis of the attack, we argue that defense mechanisms should not remain only at the level of the entire model, but should also consider the speech tokenizer. 
To this end, we propose a \textbf{RO}bust \textbf{S}p\textbf{E}ech \textbf{Tok}enizer (\textbf{ROSETok}) that achieves high-fidelity reconstruction while preserving semantically rich representations with downstream SpeechLLM against Gaussian noise and adversarial attacks.

To evaluate the effectiveness of the T-SemAttack, we conduct experiments on bilingual and multi-speaker datasets, 9 ASR systems, and 4 LALM systems, demonstrating that T-SemAttack achieves state-of-the-art (SOTA) performance in terms of perceptual quality, attack efficiency, and effectiveness. In addition, we verify the robustness of T-SemAttack against adaptive defenses, including denoising and data transformations. 
Finally, we experimentally assess the reconstruction and downstream semantic capabilities of the ROSETok under different noise conditions. Overall, the contributions of this paper can be summarized as follows:
\begin{itemize}[leftmargin=*]
    \item We reveal the vulnerability of speech tokenizers and propose a transferable semantic attack method.
    \item We explore the effectiveness and transferability of our attack from the perspectives of token vulnerability, hierarchical representation drift, and LALM attention interference.
    \item Motivated by findings of the attack causes, we propose ROSETok, a speech tokenizer that improves noise-robust reconstruction and semantic preservation, enhancing the speech systems' robustness.
    \item We conduct extensive experiments to verify the efficiency, performance, and robustness of T-SemAttack. We verify the ROSETok's performance in reconstruction and downstream scenarios.
\end{itemize}

\section{Background and Related Work}
\noindent\textbf{Speech Tokenizer and Speech Foundation Model.} A speech tokenizer typically consists of an encoder and quantizer (decoder)~\cite{codec-survey}. The former encodes speech signals into representations, while the latter discretizes these representations into specific tokens, e.g., VQ- or FSQ-based tokens~\cite{cosyvoice, cosyvoice2}. As a speech front end, the speech tokenizer plays a crucial role in modern speech foundation models and has been widely adopted in various tasks and systems, including Qwen3-ASR~\cite{qwen3-asr} for ASR, CosyVoice~\cite{cosyvoice} for TTS, and Qwen2-Audio~\cite{qwen2-audio} for LALMs. Since the speech tokenizer bridges the continuous acoustic space and the discrete semantic space, {\it \uline{safety vulnerability in this component may directly compromise the reliability and safety of the entire speech foundation model}}.

\noindent\textbf{Transferable ASR Attacks.} 
Recent adversarial attacks on ASR systems have improved transferability and input-agnostic perturbation generation~\cite{audioshield}. However, most existing methods attack the entire ASR pipeline, often relying on surrogate decoding objectives such as CTC loss~\cite{e2e-vguard}, which can incur high optimization cost and obscure the role of front-end speech representations. 
Although recent work has observed instability in speech tokenizers~\cite{stabletoken}, tokenizer-level transfer attacks and their propagation mechanisms remain underexplored. In contrast, we focus on the semantic encoder within the speech tokenizer and provide a systematic analysis of how small waveform perturbations induce semantic representation drift, token-level collapse, and downstream LALM attention disruption.

\section{Transferable Semantic Attack}\label{sec:attack_design}
\begin{figure*}[t]
    \centerline{
    \includegraphics[width=0.98\textwidth]{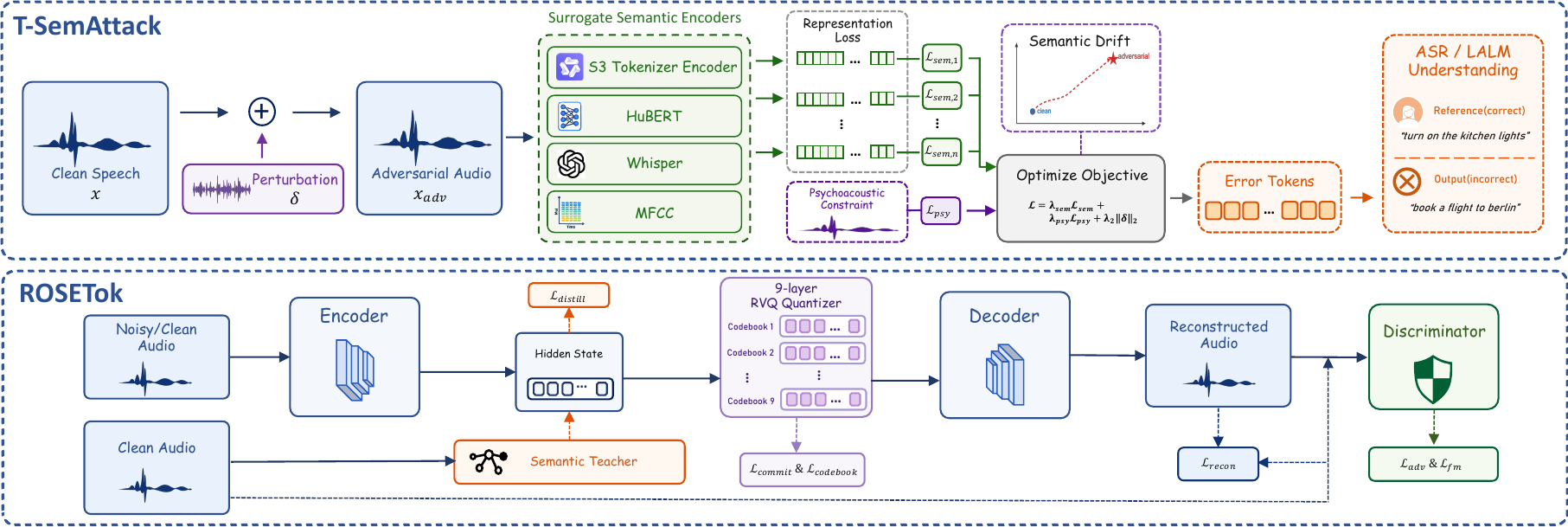}}
    \caption{The architecture and pipeline of T-SemAttack and ROSETok.}
    \label{fig_workflow}
\end{figure*}

Most existing attacks~\cite{audioshield, audiojailbreak} against speech systems rely on specific output targets of ASR or LALM models and require access to the overall architecture, which limits efficiency and transferability. As the front end of LLM-based speech systems, the speech tokenizer plays a crucial role in providing modeling inputs for LLMs, making its security particularly important. We propose T-SemAttack, which attacks semantic representations by jointly perturbing surrogate semantic encoders and uses psychoacoustic constraints to enhance the perceptual quality in Figure \ref{fig_workflow}.

\subsection{Threat Model}
\noindent\textbf{Adversary Assumption.} We consider a transfer-based attack scenario in which the adversary can access the input speech waveform and possesses a set of differentiable surrogate semantic encoders, but cannot access the internal gradients, parameters, or tokenizer details of the target systems. Under the constraint that the noise remains acceptable, the adversary aims to generate robust adversarial examples that cause speech systems to produce incorrect transcriptions. The adversarial examples should also be robust against adaptive defense, e.g., denoising and data transformations. Moreover, our attack is conducted in an offline setting with minimal real-time requirements.

\noindent\textbf{Defense Assumption.} We assume a strong defender who can detect the embedded noise and can use existing relevant tools or defense strategies to restore the original audio as much as possible.

\subsection{Problem Definition}
\begin{tcolorbox}
  \begin{center}
    \textit{Can perturbations disrupt the semantic representations of both continuous and discrete speech front ends \textbf{in a unified manner}, thereby affecting downstream speech models?}
  \end{center}
\end{tcolorbox}

Both continuous and discrete speech tokenizers typically produce continuous representations, with discrete tokenizers additionally applying quantization. Hard quantization is locally constant within quantization regions and non-differentiable at their boundaries. \textit{We therefore target the continuous representations before any optional quantization}. Representation drift can alter the features supplied through a continuous interface or change discrete token assignments when the perturbed representations cross quantization boundaries. We evaluate the downstream effects empirically.

Given an original audio signal $x \in [-1,1]^T$, where $T$ denotes the audio length, our goal is to construct an adversarial example $x_{\mathrm{adv}} = x + \delta$, where $\delta$ is a small adversarial perturbation. The overall optimization objective can be written as Eq. (\ref{eq:definition}).
\begin{equation}
    \begin{aligned}
    & \min_{\delta} \quad \mathcal{L} =  \lambda_{\mathrm{sem}} \mathcal{L}_{\mathrm{sem}}(x, x_{\mathrm{adv}}) + \lambda_{\mathrm{psy}} \mathcal{L}_{\mathrm{psy}}(x,\delta) + \lambda_{2}\|\delta\|_2, \\
    & \text{s.t.} \quad \|\delta\|_\infty \le \epsilon,  \quad x_{\mathrm{adv}} = x + \delta \in [-1,1]^T,
    \end{aligned}
    \label{eq:definition}
\end{equation}
where $\mathcal{L}_{\mathrm{sem}}$ measures the deviation of semantic representations, $\mathcal{L}_{\mathrm{psy}}$ represents psychoacoustic masking, and $\|\delta\|_2$ controls perturbation level. We use $\lambda_{\mathrm{sem}}=100$, $\lambda_{\mathrm{psy}}=10^{-3}$, and $\lambda_2=0.1$.

\subsection{Tokenizer-centric Semantic Attack}
\noindent\textbf{Representation Drifting.} Representation drift refers to a shift in representations of perturbed speech relative to those of clean speech. T-SemAttack optimizes waveform perturbations to reduce the cosine similarity between these representations in the encoding space, causing different internal representations and thereby disrupting downstream models’ understanding of speech content.

To improve cross-model transferability, we jointly optimize over multiple encoders. Specifically, we construct the following four surrogate branches: (1) \textbf{S3Tokenizer}~\cite{cosyvoice}, which directly simulates the semantic encoding process of the \textit{discrete} speech tokenizer. (2) \textbf{HuBERT}~\cite{hubert}, which provides deep semantic representations obtained through self-supervised learning. (3) \textbf{Whisper} encoder~\cite{whisper}, which provides robust semantic representations learned from strong ASR pretraining. (4) Mel-Frequency Cepstral Coefficient (\textbf{MFCC})~\cite{advddos, e2e-vguard}, is used as an additional low-level attack target~\cite{e2e-vguard, zq-attack}.

Let $\mathcal{F}={f_1,\dots,f_K}$ denote the $K$ surrogate semantic encoders. For input $x$, encoder $f_k$ outputs representations $f_k(x)$. Eq. (\ref{eq:loss_sem}) defines the semantic loss from frame-wise averaged cosine similarity.
\begin{equation}
    \mathcal{L}_{\mathrm{sem}} = \sum_{k=1}^{K}  \operatorname{CosSim}\big(f_k(x), f_k(x_{\mathrm{adv}})\big).
    \label{eq:loss_sem}
\end{equation}
By minimizing Eq. (\ref{eq:loss_sem}), the attack simultaneously drives shifts in multiple semantic spaces, thereby improving transferability to unseen ASR and LALM models.

Mechanistically, this attack can transfer to different ASR and LALM models because multiple semantic encoders, although trained with different objectives, share similar Mel-input bottlenecks and semantic structures. Therefore, disrupting the shared semantic representations in surrogate models often also disrupts the corresponding representations in unseen target models.

\noindent\textbf{Psychoacoustic Model.} Previous transferable ASR attacks suffer from low perceptual quality. We utilize a psychoacoustic constraint to ensure that the adversarial perturbation remains acceptable to humans.
Given the original audio $x$, we first compute the corresponding masking threshold $\Theta(x)$ based on its short-time spectrum and use it as an upper-bound reference for the perturbation energy. Let $\operatorname{PSD}(\delta)$ denote the power spectral density of the perturbation in the time-frequency domain. The psychoacoustic loss can then be written as Eq. (\ref{eq:loss_psy}).
\begin{equation}
    \mathcal{L}_{\mathrm{psy}} = \sum_{t,f} \max\big(0, \operatorname{PSD}_{t,f}(\delta) - \Theta_{t,f}(x) \big),
    \label{eq:loss_psy}
\end{equation}
which imposes a penalty only when the perturbation energy in a specific time-frequency unit exceeds the masking threshold. In this way, the optimization process preferentially hides the perturbation in frequency bands or local temporal regions where the original speech is stronger, thereby balancing attack effectiveness and perceptual quality. We follow prior work~\cite{voicecloak}.

\noindent\textbf{Optimization Process.}
In optimization, we extract multiple representations and update the adversarial audio based on the total loss in Eq. (\ref{eq:definition}). The optimization adopts a PGD-style~\cite{pgd} projected update, where the perturbation is projected back onto the $\ell_\infty$ ball after each step in Eq. (\ref{eq:pgd}).
\begin{equation}
    \delta \leftarrow \operatorname{Proj}_{\|\delta\|_\infty \le \epsilon} \left( \delta - \alpha \cdot  \operatorname{sign} \left( \nabla_\delta \mathcal{L} \right) \right),
    \label{eq:pgd}
\end{equation}
where $\alpha$ is the step size. In practice, we combine the SGD optimizer with an adaptive plateau learning-rate scheduler~\cite{e2e-vguard}, which automatically reduces the learning rate when the loss stagnates, thereby improving optimization stability. The entire process is run for 500 iterations.

\section{Why does the T-SemAttack work?}\label{sec:explanation}
In Section \ref{sec:experiments}, we conduct extensive experiments demonstrating the effectiveness of T-SemAttack. In this section, we explore why T-SemAttack is effective and how semantic attacks affect the model.

\begin{figure}[t]
    \centering
    \begin{subfigure}[t]{0.33\linewidth}
        \centering
        \includegraphics[width=\linewidth]{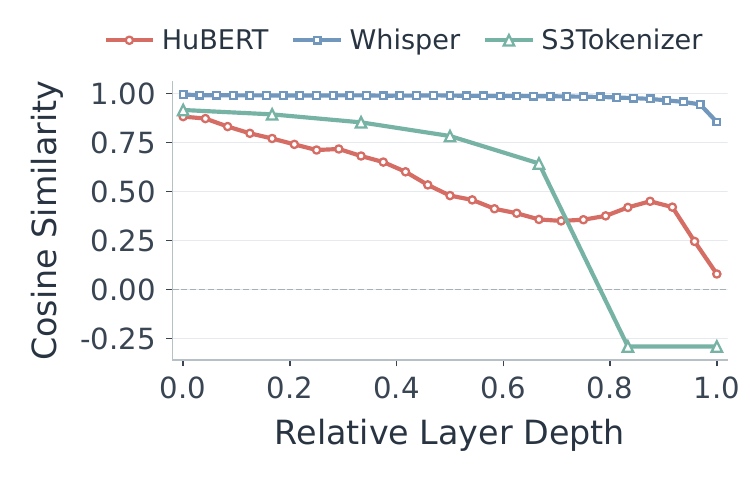}
        \caption{Layer-wise similarity.}
        \label{fig:cosine_similarity}
    \end{subfigure}
    \hfill
    \begin{subfigure}[t]{0.66\linewidth}
        \centering
        \includegraphics[width=\linewidth]{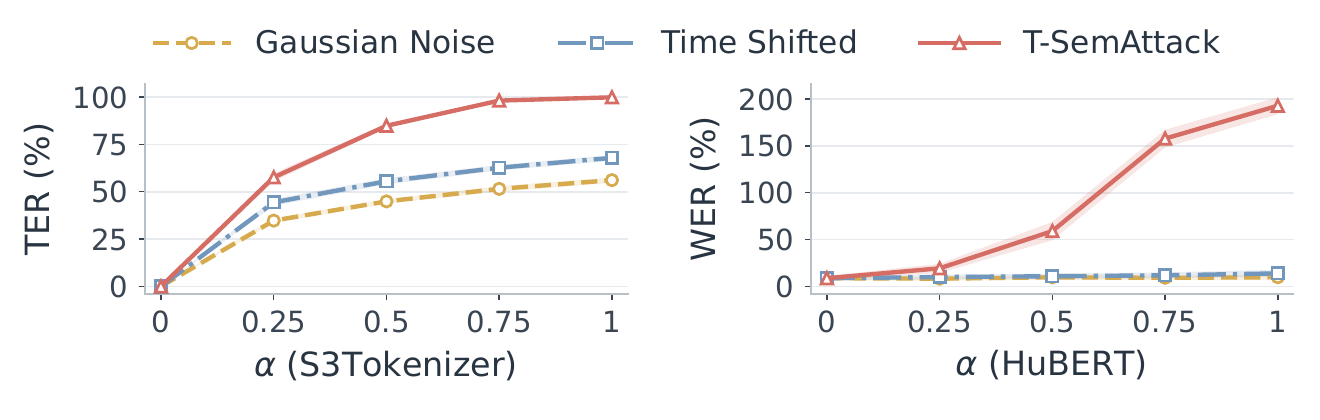}
        \caption{Effect of perturbation dose controlled by $\alpha$.}
        \label{fig:perturbation_dose}
    \end{subfigure}
    \caption{Layer-wise analysis of representation similarity and effectiveness of perturbation magnitude.}
    \label{fig:layerwise_analysis}
\end{figure}

\noindent\textbf{Layer-wise Representation Similarity.} We compare the layer-wise representations of clean and attacked speech on LibriTTS. As shown in Figure \ref{fig:cosine_similarity}, HuBERT’s cosine similarity decreases from 0.881 at the input embedding layer to 0.079 at the final layer. Whisper’s shallow representations remain relatively stable, but its final-layer similarity still drops to 0.854. These results indicate that \textit{small waveform perturbations can induce substantial shifts in deep representations}, highlighting the need for defenses that promote semantic stability throughout the encoding process.

\noindent\textbf{Shared Attack Space.} We extract a candidate shared subspace via singular value decomposition of normalized gradients and compare different attack directions under the same perturbation budget. Across 30 samples, using only the rank-1 shared direction yields a WER of 155.02\% on the model HuBERT, comparable with the 191.64\% achieved by the full ensemble attack and the 26.38\% achieved by the residual attack. These findings suggest that shared directions can support strong cross-model transferability, while the contributions of different attack components depend on the target model.

\noindent\textbf{Perturbation Magnitude.} We test the attack direction and scale the perturbation according to $x_\alpha = x + \alpha\delta$, comparing it with energy-matched Gaussian noise and temporally shifted perturbations. As $\alpha$ increases from 0.5 to 0.75, HuBERT’s WER rises from 59.05\% to 157.74\%, reaching 192.84\% at $\alpha = 1$, compared with only 9.83\% and 13.92\% for the two controls, respectively. Meanwhile, at $\alpha = 1$, the token error rate (TER) of S3Tokenizer reaches 99.88\%, exceeding the 56.12\% observed for the Gaussian noise, respectively. These results indicate that \textit{an appropriate direction can yield stronger attacks with less perturbation energy}.

\section{Mechanism-Guided Robust Speech Tokenizer}
In Section \ref{sec:explanation}, we investigate why T-SemAttack can achieve effective attacks and reveal an important insight: \textit{some existing speech tokenizers are not semantically robust and are vulnerable to adversarial attacks. We believe that defense solely at the LLM level is more time- and cost-intensive.} Therefore, we return to the core of the problem and aim to improve the stability of the tokenizer against noise. In this work, we utilize noisy training and semantic distillation to improve the robustness.

\textbf{Semantic Preservation Module.} 
Figure \ref{fig:cosine_similarity} illustrates the loss of deep semantic information. We employ a pretrained Semantic Teacher, Whisper~\cite{whisper}, to supervise the representations of the model encoder. Specifically, we align the representations of noisy inputs with those of clean audio, thereby preserving semantic information and improving robustness to noise in downstream tasks.

\textbf{Two-Stage Training.} We utilize a two-stage training strategy consisting of base pretraining followed by attack-aware fine-tuning. In the first stage, we train on English and Chinese speech from Emilia, using speech corrupted by Gaussian noise as input and the corresponding clean audio as the reconstruction target, together with semantic distillation. We use a learning rate of $1 \times 10^{-4}$ to achieve high-fidelity reconstruction, semantic preservation, and general robustness to noise, yielding \textbf{ROSETok-Base}. In the second stage, to further adapt the model to structured semantic perturbations, we randomly select 2,000 utterances from LibriSpeech and generate perturbed inputs using T-SemAttack. We then perform Attack-Aware Fine-Tuning (AAF) on ROSETok-Base, using the original clean audio and its teacher representations as supervision. We reduce the learning rate to $5 \times 10^{-6}$ to preserve the capabilities acquired during pretraining while improving robustness to adversarial perturbations, yielding \textbf{ROSETok-AAF}. Our model architecture is based on DAC and consists of an encoder, a nine-layer quantizer, and a decoder.

\textbf{Training Objectives.} We aim for high-fidelity reconstruction and semantic consistency. For both clean and noisy inputs, the model outputs high-quality clean audio. Moreover, we hope the encoder can preserve semantically rich information even for noisy audio inputs. We use a multi-frequency discriminator with hinge loss~\cite{llasa} for adversarial learning. 
The objective is in Eq. (\ref{eq:train}).
\begin{equation}
    \mathcal{L}_{\mathrm{tok}} = \lambda_{\mathrm{distill}} \mathcal{L}_{\mathrm{distill}} + \lambda_{\mathrm{recon}} \mathcal{L}_{\mathrm{recon}} + \lambda_{\mathrm{cb}} \mathcal{L}_{\mathrm{cb}} + \lambda_{\mathrm{commit}} \mathcal{L}_{\mathrm{commit}} + \mathcal{L}_{\mathrm{adv}} + \mathcal{L}_{\mathrm{fm}},
    \label{eq:train}
\end{equation}
where $\mathcal{L}_{\mathrm{distill}}$ is the semantic distillation loss, $\mathcal{L}_{\mathrm{recon}}$ is the reconstruction loss, $\mathcal{L}_{\mathrm{cb}}$ is the codebook loss, $\mathcal{L}_{\mathrm{commit}}$ is the commitment loss, $\mathcal{L}_{\mathrm{adv}}$ is the adversarial loss, and $\mathcal{L}_{\mathrm{fm}}$ is the feature matching loss. Detailed descriptions are provided in Appendix \ref{sec:appendix_details_training}. The hyperparameters are set as
$\{\lambda_{\mathrm{distill}}, \lambda_{\mathrm{recon}}, \lambda_{\mathrm{cb}}, \lambda_{\mathrm{commit}}\} = \{5.0, 15, 1.0, 0.25\}$. The weights of $\mathcal{L}_{\mathrm{adv}}$ and $\mathcal{L}_{\mathrm{fm}}$ are both 1.

\section{Experimental Setup}\label{sec:experimental_setup}
\subsection{Semantic Attack Setup}
\noindent\textbf{Datasets.} To comprehensively evaluate the effectiveness of the semantic attack, we use bilingual and multi-speaker speech datasets. LibriTTS~\cite{libritts}: Following \cite{e2e-vguard}, we select single-speaker samples to evaluate the effectiveness of the attack in an English single-speaker scenario. Seed-TTS-Eval~\cite{seedtts}: This is a widely used benchmark, containing both Chinese and English utterances from multiple speakers. We randomly sample 200 English utterances and 200 Chinese utterances for evaluation.

\noindent\textbf{ASR Systems.} We evaluate the attack performance on both open-source models (Qwen3-ASR~\cite{qwen3-asr}, VibeVoice-ASR~\cite{vibevoice-asr}, FireRed-ASR~\cite{fireredasr}, and Fun-ASR~\cite{fun-asr}) and commercial black-box models (iFlytek, Doubao ASR, and GPT-4o-mini-transcribe).

\noindent\textbf{LALM Systems.} The safety of LALMs is also a major focus of this work. We further evaluate our attack on several representative LALMs, including the open-source models Qwen2-Audio~\cite{qwen2-audio}, Qwen3-Omni~\cite{qwen3-omni}, and Audio-Flamingo 3~\cite{af3}, as well as the black-box model Gemini-3-Flash.

\noindent\textbf{Metrics.} For efficiency, we report the Real-Time Factor (RTF) to measure the computational cost of the attack. For perceptual quality, we use Signal-to-Noise Ratio (SNR) and Mean Opinion Score (MOS) to evaluate the degree of distortion introduced by the attack in terms of audio quality.
For attack effectiveness, we transcribe the adversarial audio using the target model and compute the Word Error Rate (WER). A higher WER indicates more severe corruption of the semantic content in the original speech.

\subsection{Training Details of the Robust Tokenizer}
\noindent\textbf{Datasets.} We train the ROSETok-Base on approximately 96.7k hours of bilingual speech data in Chinese and English. We utilize the English and Chinese subset from the Emilia~\cite{emilia} dataset.

\noindent\textbf{Training Hyperparameters.} We train the model on 8 NVIDIA H100 GPUs with a global batch size of 128. We use the AdamW optimizer, with an initial learning rate of $1 \times 10^{-4}$ and $\beta_1 = 0.8, \beta_2 = 0.99$. The model is trained for a total of 200k steps, and the learning rate is decayed by 10\%.

\section{Experiments}\label{sec:experiments}
\textbf{Organization.} First, we verify the effect and transferability of attacks on ASR systems. Second, we evaluate the impact of the attack on cross-modal audio understanding in LALMs. Third, we test the adaptive robustness of the attack against denoising, adversarial defense, and audio transformation techniques. For ROSETok, we evaluate its reconstruction stability on both clean and noisy audio. Finally, we conduct ablation studies for validation. Case study is provided in Appendix \ref{sec:appendix_case_study}.

\subsection{Evaluation on ASR Systems}

In Table \ref{tab:asr_libritts}, T-SemAttack achieves a better balance among efficiency, perceptual quality, and attack strength. Its RTF is 21.800, lower than E2E-VGuard's 26.797, while maintaining an SNR of 16.319 with good perceptual quality. Subjective perceptual experiments provided in Appendix \ref{sec:human_study} show T-SemAttack achieve human acceptable perception. Details and results of AudioShield's efficiency comparison are provided in Appendices \ref{sec:discussion} and \ref{sec:appendix_cost}. AudioShield can produce relatively high WERs, but it almost destroys the audio quality, making the content difficult to hear clearly (Table \ref{tab:human_study}). In contrast, with better perceptual quality, T-SemAttack increases the WERs of multiple open-source and black-box ASR systems from about 1\%$\sim$4\% on clean speech to 20.004\%$\sim$63.555\%, reaching 49.607\% and 46.089\% on iFlytek and GPT-4o-mini-transcribe, respectively.

Table \ref{tab:asr_libritts} verifies generalization in multi-speaker and bilingual scenarios. On Seed-TTS (EN), T-SemAttack achieves an average WER of 81.872\%, higher than E2E-VGuard's 23.154\%. On the Chinese subset, the average WER remains 50.814\%. Notably, T-SemAttack achieves a cross-linguistic attack on the Chinese dataset because our surrogate model, (Whisper and S3Tokenizer) are trained on large-scale multilingual datasets. Commercial systems also show robustness differences in Chinese scenarios. This experiment indicates that the transferability of semantic attacks generally holds.

\begin{table}[t]
      \centering
      \caption{Comparison of attack efficiency, perceptual quality, and attack effectiveness across ASR systems and datasets. WER (\%) is reported, where higher values indicate stronger attack effectiveness. The best results are highlighted and second-best results are \uline{underlined}.}
      \label{tab:asr_libritts}
      \resizebox{\textwidth}{!}{%
      \begin{tabular}{l*{12}{c}}
      \toprule
      \multirow{2}{*}{\textbf{Methods}}
      & \multirow{2}{*}{\textbf{RTF}($\downarrow$)}
      & \multirow{2}{*}{\textbf{SNR}($\uparrow$)}
      & \multicolumn{2}{c}{\textbf{Surrogate Model}}
      & \multicolumn{4}{c}{\textbf{Open-source Model}}
      & \multicolumn{3}{c}{\textbf{Black-box Model}}
      & \multirow{2}{*}{\textbf{Avg.}} \\
      \cmidrule(lr){4-5}
      \cmidrule(lr){6-9}
      \cmidrule(lr){10-12}
      & &
      & HuBERT & Whisper
      & Qwen3-ASR & VibeVoice
      & FireRed-ASR & Fun-ASR
      & iFlytek & Doubao & GPT-4o & \\
      \midrule

      \multicolumn{13}{l}{\textit{LibriTTS: A single-speaker English dataset.}} \\

      Clean
      & -- & --
      & 2.694 & 2.079
      & 2.369 & 2.156 & 2.017 & 1.866
      & 3.648 & 3.463 & 2.234 & 2.503 \\

      Gaussian Noise
      & 0.003 & 15.197
      & 3.228 & 2.213
      & 2.583 & 2.736 & 2.933 & 2.488
      & 3.837 & 3.691 & 6.677 & 3.376 \\

      E2E-VGuard
      & 26.797 & 14.341
      & 10.781 & 5.834
      & 6.781 & 8.325 & 11.376 & 8.842
      & 8.672 & 7.618 & 8.030 & 8.473 \\

      AudioShield
      & -- & -2.440
      & 16.007 & 10.363
      & 15.450 & 15.241 & 13.800 & 12.171
      & 16.815 & 14.273 & 11.049 & 13.908 \\

      \textbf{T-SemAttack}
      & 21.800 & \best{16.319}
      & \best{156.913} & \best{111.367}
      & \best{27.388} & \best{63.555}
      & \best{33.529} & \best{43.679}
      & \best{49.607} & \best{20.004}
      & \best{46.089} & \best{61.348} \\

      \midrule
      \multicolumn{13}{l}{\textit{Seed-TTS (EN): A multi-speaker English dataset.}} \\

      Clean
      & -- & --
      & 6.286 & 0.851
      & 1.040 & 1.433 & 5.906 & 1.015
      & 1.942 & 3.001 & 1.907 & 2.598 \\

      Gaussian Noise
      & 0.003 & \uline{15.228}
      & 10.739 & 1.307
      & 1.695 & 3.862 & 10.985 & 2.061
      & 3.457 & 2.679 & 9.849 & 5.182 \\

      E2E-VGuard
      & 26.797 & 12.907
      & 42.084 & 15.913
      & 15.643 & 26.609 & 35.936 & 20.541
      & 19.051 & 14.233 & 18.374 & 23.154 \\

      AudioShield
      & -- & -2.364
      & \uline{62.779} & \uline{42.840}
      & \best{50.674} & \uline{51.296}
      & \uline{54.301} & \uline{56.806}
      & \uline{54.181} & \best{61.297}
      & \uline{42.172} & \uline{52.927} \\

      \textbf{T-SemAttack}
      & 21.800 & \best{17.598}
      & \best{189.289} & \best{125.913}
      & \uline{37.445} & \best{103.887}
      & \best{65.406} & \best{63.779}
      & \best{71.726} & \uline{29.948}
      & \best{49.457} & \best{81.872} \\

      \midrule
      \multicolumn{13}{l}{\textit{Seed-TTS (ZH): A multi-speaker Chinese dataset.}} \\

      Clean
      & -- & --
      & -- & 3.086
      & 0.790 & 1.521 & 1.760 & 0.614
      & 1.560 & 1.109 & 2.984 & 1.678 \\

      Gaussian Noise
      & 0.003 & \uline{15.387}
      & -- & 5.434
      & 1.207 & 6.171 & 2.899 & 1.466
      & 1.331 & 0.774 & 15.129 & 4.301 \\

      E2E-VGuard
      & 26.797 & 13.607
      & -- & 18.392
      & 5.322 & 27.258 & 8.896 & 6.879
      & 3.119 & 1.944 & 18.279 & 11.261 \\

      AudioShield
      & -- & -2.435
      & -- & \uline{28.113}
      & \best{19.998} & \uline{50.729}
      & \uline{16.609} & \uline{23.267}
      & \uline{14.345} & \best{30.221}
      & \uline{37.687} & \uline{27.621} \\

      \textbf{T-SemAttack}
      & 21.800 & \best{17.652}
      & -- & \best{109.100}
      & \uline{18.833} & \best{144.637}
      & \best{19.293} & \best{29.021}
      & \best{19.214} & \uline{6.298}
      & \best{60.116} & \best{50.814} \\

      \bottomrule
      \end{tabular}%
      }
\end{table}

\subsection{Evaluation on LALMs}
\begin{figure*}[t]
  \centering
  \includegraphics[width=0.9\textwidth]{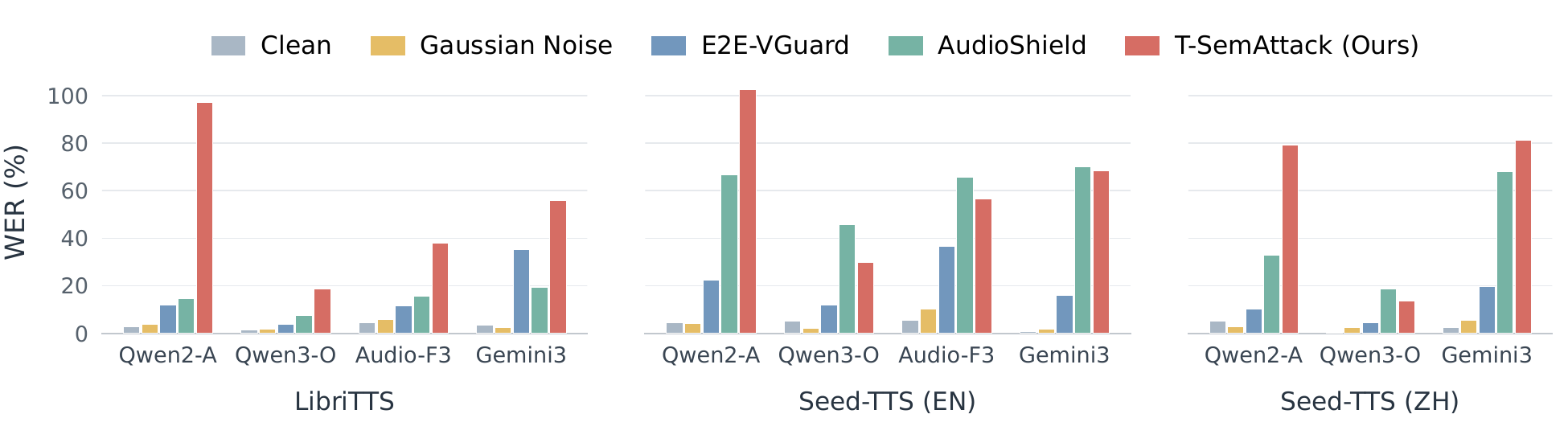}
  \caption{ASR performance across different datasets and LALMs, measured by WER (\%).}
  \label{fig:wer_comparison}
\end{figure*}

For evaluation on LALMs, we consider two tasks, i.e., ASR and question-answering tasks.

\noindent\textbf{ASR Evaluation.} Figure \ref{fig:wer_comparison} shows the impact of T-SemAttack on the speech understanding ability of LALMs. Compared with ASR systems, LALMs not only need to recognize speech content, but also {\it align audio tokens with the text reasoning process}. Therefore, they are more sensitive to semantic drift at the tokenizer level. T-SemAttack achieves an average WER of 58.30\% over 11 dataset-model combinations, which is much higher than E2E-VGuard's 16.76\% and AudioShield's 38.66\%. For example, on LibriTTS, the WER of Qwen2-Audio increases from 2.982\% to 97.280\%. On the Chinese set, Qwen2-Audio and Gemini3-Flash also reach 79.259\% and 81.300\%, respectively. These results show that once the semantic tokenizer produces incorrect semantic tokens, subsequent LALMs can hardly recover the original semantics through stronger language modeling. 

This phenomenon is also consistent with the attention analysis in Section \ref{sec:lalm_attn} which shows that attacked audio tokens weaken the model's reliance on audio context, causing cross-modal attention to shift more toward the text itself and thereby amplifying semantic understanding errors.

\noindent\textbf{QA Evaluation.} we evaluate four LALMs on the \textit{audio question-answering (QA)} task on the AIR-Bench~\cite{air-bench} with randomly selected 100 samples. We use DeepSeek-V4-Flash as the judge model to assign a Response Quality Score (RQS), where higher scores indicate better response quality. As shown in Table~\ref{tab:airbench_qa}, all four LALMs receive lower RQS values under T-SemAttack, with the mean across models decreasing from 7.58 on clean inputs to 5.48 on perturbed inputs. These results show that T-SemAttack can also degrade the quality of LALMs' audio QA performance.

\begin{table*}[t]
      \centering
      \caption{\textbf{Robustness evaluation} against advanced denoising models across two models, adversarial
      audio defense, and audio transformations on the VibeVoice-ASR with averaged WER(\%) values.}
      \label{tab:defense_transformation}
      \resizebox{\textwidth}{!}{
      \begin{tabular}{lcccccccccccccc}
      \toprule
      \multirow{2}{*}{\textbf{Method}}
      & \multicolumn{2}{c}{\textbf{Denoiser}}
      & \multicolumn{2}{c}{\textbf{DeepFilterNet3}}
      & \multicolumn{3}{c}{\textbf{Adversarial Defense}}
      & \multicolumn{1}{c}{\textbf{Diffusion}}
      & \multicolumn{3}{c}{\textbf{Audio Transformation}} \\
      \cmidrule(lr){2-3}
      \cmidrule(lr){4-5}
      \cmidrule(lr){6-8}
      \cmidrule(lr){9-9}
      \cmidrule(lr){10-12}
      & Whisper & VibeVoice & Whisper & VibeVoice
      & Resample & Mel & Q-D
      & AudioPure
      & Speed & PS & Band-filter \\
      \midrule

      Gaussian Noise
          & 2.964 & 1.720
          & 2.468 & 1.133
          & 5.256 & 6.569 & 2.984
          & 8.737
          & 57.517 & 3.735 & 20.334 \\

      E2E-VGuard
          & 10.205 & 7.276
          & 10.168 & 7.148
          & 13.841 & 17.650 & 9.447
          & 13.898
          & 67.112 & 10.282 & 28.863 \\
          
      AudioShield
          & 9.930 & 9.738
          & 10.741 & 9.187
          & 16.030 & 16.751 & 14.673
          & \textbf{37.497}
          & 61.546 & 13.379 & 23.901 \\

      \textbf{T-SemAttack}
          & \textbf{96.763} & \textbf{42.901}
          & \textbf{113.997} & \textbf{37.805}
          & \textbf{55.805} & \textbf{58.997} & \textbf{73.364}
          & 21.571
          & \textbf{89.898} & \textbf{70.635} & \textbf{71.085} \\

      \bottomrule
      \end{tabular}
      }
  \end{table*}

\begin{table}[htbp]
    \centering
    \setlength{\tabcolsep}{4pt}
    \renewcommand{\arraystretch}{1.1}
    \caption{Response Quality Scores of LALMs on AIR-Bench with and without T-SemAttack. Avg. denotes the mean across the four models.}
    \label{tab:airbench_qa}
    \resizebox{0.8\textwidth}{!}{
    \begin{tabular}{@{}lccccc@{}}
        \toprule
        \textbf{Input}
        & \textbf{Qwen2-Audio}
        & \textbf{Qwen3-Omni}
        & \textbf{Audio-Flamingo 3}
        & \textbf{Gemini 3 Flash}
        & \textbf{Avg.} ($\downarrow$) \\
        \midrule
        Clean
        & 6.63 & 7.73 & 8.14 & 7.81 & 7.58 \\
        \textbf{T-SemAttack}
        & \best{4.58} & \best{5.23} & \best{6.20} & \best{5.89} & \best{5.48} \\
        \bottomrule
    \end{tabular}
    }
\end{table}

\subsection{Adaptive Defense against Semantic Attack}

We consider some strong defense settings: denoising, adversarial defense, diffusion-based adversarial defense, and audio transformations. This experiment is conducted on the Seed-TTS-EN dataset.

\textbf{Advanced Denoising.} We utilize two SOTA denoising models, Denoiser~\cite{denoiser} and DeepFilterNet3~\cite{deepfilternet3}. The results in Table \ref{tab:defense_transformation} show that after Denoiser processing, T-SemAttack still achieves WERs of 96.763\% and 42.901\% on Whisper, and VibeVoice-ASR, respectively. The WERs remain high after DeepFilterNet3 processing. In contrast, under the same defenses, the WERs of E2E-VGuard and AudioShield are mostly only 7\%$\sim$16\%.

\textbf{Adversarial Defense and Audio Transformations.} 
We consider adversarial defense based on WaveGuard, diffusion-based defense (AudioPure~\cite{audiopure}), and audio transformation-based defense. The results are shown in Table \ref{tab:defense_transformation}. The best defense method is AudioPure, achieving a WER score of 21.571\%, and still misleads correct transcription. The highest WER of AudioPure-perturbed samples is 91.667\%, reflecting that this method cannot perfectly disrupt the T-SemAttack.
We find that previous methods are not robust enough against these advanced defensive methods, and T-SemAttack achieves SOTA performance. 
The explanations of these techniques are provided in the Appendix \ref{appendix:details_adv_defense}.

These results suggest that T-SemAttack perturbations are not merely removable noise by conventional speech enhancement. Although Denoiser and DeepFilterNet3 can suppress the acoustic disturbance, they cannot reliably recover the original token sequence or semantic representation once the tokenizer output has been shifted. This indicates that speech enhancement alone is insufficient, and {\it ROSETok must explicitly model both noise robustness and semantic consistency at the tokenizer level.}

\subsection{Robust Speech Tokenizer Analysis}

\begin{table*}[t]
    \centering
    \caption{Reconstruction and downstream evaluation of speech tokenizers under different inputs.}
    \label{tab:tokenizer_robustness}
    \resizebox{0.8\textwidth}{!}{
    \begin{tabular}{lcccccccc}
    \toprule
    \multirow{2}{*}{\textbf{Model}}
    & \multirow{2}{*}{\textbf{$N_q$}}
    & \multirow{2}{*}{\textbf{RTF}($\downarrow$)}
    & \multicolumn{3}{c}{\textbf{Reconstruction}($\uparrow$)}
    & \multicolumn{3}{c}{\textbf{Downstream SpeechLLM}($\uparrow$)} \\
    \cmidrule(lr){4-6}
    \cmidrule(lr){7-9}
    & & & Clean & Gaussian & T-SemAttack
      & Clean & Gaussian & T-SemAttack \\
    \midrule
    
    \multicolumn{8}{l}{\textit{Acoustic and Semantic Speech Tokenizer}} \\
    SNAC
        & 4 & 0.003
        & 0.874 & 0.817 & 0.782
        & 28.3 & 5.6 & 14.6 \\
    XY-Tokenizer
        & 8 & 0.025
        & 0.905 & 0.849 & 0.762
        & 92.3 & 87.8 & 23.1 \\
    X-Codec 2
        & 1 & 0.021
        & 0.864 & 0.808 & 0.771
        & 20.1 & 8.1 & 7.4 \\
    DAC
        & 9 & 0.008
        & 0.943 & 0.891 & 0.830
        & 25.1 & 9.1 & 13.9 \\
    \midrule
    
    \multicolumn{8}{l}{\textit{Robust Speech Tokenizer}} \\
    MOSS-Audio-Tokenizer
        & 9 & 0.029
        & 0.935 & 0.894 & 0.831
        & 54.1 & 27.6 & 27.4 \\
        
    StableToken (Pre-Q)
        & \multirow{2}{*}{1} & \multirow{2}{*}{0.175}
        & \multirow{2}{*}{0.647} 
        & \multirow{2}{*}{0.647} 
        & \multirow{2}{*}{0.592}
        & \textbf{96.7} & \textbf{95.4} & 35.0 \\

    StableToken (Post-Q)
        & &
        & & &
        & 91.0 & 89.5 & 27.9 \\
    \textbf{ROSETok-Base}
        & \multirow{2}{*}{9} & \textbf{0.010}
        & \textbf{0.966} & \textbf{0.950} & 0.842
        & 93.0 & 92.9 & 62.4 \\
    \textbf{ROSETok-AAF}
        & & --
        & 0.945 & 0.937 & \best{0.903}
        & 92.7 & 89.3 & \best{77.9} \\
    \bottomrule
    \end{tabular}
    }
    \par\vspace{4pt}
    \begin{minipage}{\linewidth}
        \footnotesize
        \textit{Note.}
        Pre-Q and Post-Q denote representations before and after quantization for downstream evaluation. All other models use Pre-Q features. Bold metric values indicate the best reported result in each column.
    \end{minipage}
\end{table*}

\noindent\textbf{Reconstruction Performance.} We first evaluate speech tokenizers' reconstruction performance on Seed-TTS (EN) using Short-Time Objective Intelligibility (STOI). As shown in Table \ref{tab:tokenizer_robustness}, ROSETok-Base achieves reconstruction scores of 0.966, 0.950, and 0.842 on clean, Gaussian-noise-corrupted, and T-SemAttack-perturbed inputs, respectively, outperforming baselines. ROSETok-AAF further improves the reconstruction score under T-SemAttack to 0.903 while maintaining high reconstruction quality on clean and Gaussian-noise-corrupted inputs. These results demonstrate that two-stage training can preserve reconstruction quality while improving adversarial robustness.

\noindent\textbf{Downstream Semantic Robustness.} Following the XARES-LLM benchmark~\cite{xares-llm}, we feed each tokenizer's pre-quantized continuous representations into SmolLM2-135M and evaluate their impact on downstream semantic understanding using the FSC task. We train the LLM with clean samples and test with different inputs. Existing tokenizers generally exhibit substantial degradation under semantic attacks. For example, StableToken achieves scores of 96.7 and 95.4 on clean and Gaussian-noise-corrupted inputs, respectively, but only 35.0 under T-SemAttack, indicating that robustness to Gaussian noise is insufficient to ensure downstream semantic understanding. In comparison, ROSETok-Base achieves 62.4 under attack, and ROSETok-AAF further improves this score to 77.9. \textit{These results show that the robustness of front-end representations affects downstream performance and that AAF can mitigate performance degradation from semantic attacks.}

\begin{figure*}[t]
    \centering

    \begin{minipage}[t]{0.42\textwidth}
        \vspace{0pt}
        \centering
        \includegraphics[width=\linewidth]{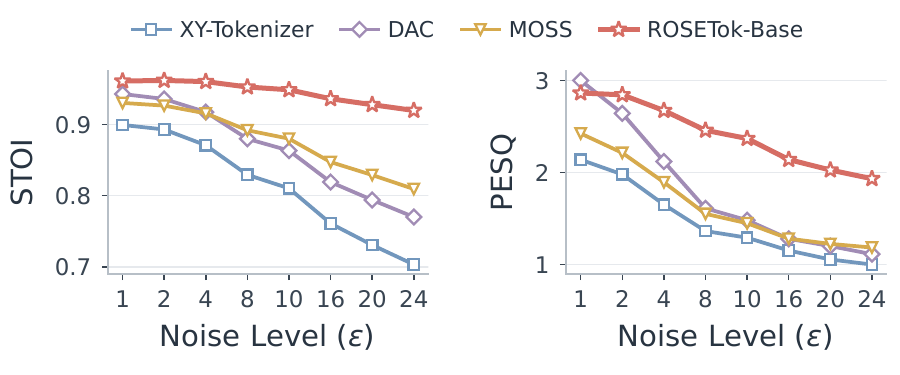}
        \captionof{figure}{Noise-level evaluation.}
        \label{fig_noise-level}
    \end{minipage}
    \hfill
    \begin{minipage}[t]{0.50\textwidth}
        \vspace{0pt}
        \centering
        \captionof{table}{Ablation results on reconstruction and semantic evaluation of the FSC task with MLP. ``w/o'' represents without this loss/component.}
        \label{tab:ablation_rosetok}
        \resizebox{\linewidth}{!}{
        \begin{tabular}{lcccccc}
        \toprule
        \multirow{2}{*}{Method} 
        & \multicolumn{2}{c}{Recon-Clean} 
        & \multicolumn{2}{c}{Recon-Noisy} 
        & \multicolumn{2}{c}{Semantic-FSC($\uparrow$)} \\
        \cmidrule(lr){2-3} \cmidrule(lr){4-5} \cmidrule(lr){6-7}
        & STOI($\uparrow$) & PESQ($\uparrow$) 
        & STOI($\uparrow$) & PESQ($\uparrow$) 
        & Clean & Noisy \\
        \midrule
        w/o noisy 
            & \textbf{0.956} & \textbf{2.892} & 0.879 & 1.578 
            & \textbf{69.259} & \uline{11.363} \\
        w/o $\mathcal{L}_{\mathrm{distill}}$ 
            & \uline{0.947} & \uline{2.652} & \textbf{0.934} & \textbf{2.322} 
            & 7.118 & 5.220 \\
        \textbf{Ours} 
            & 0.945 & 2.573 & \uline{0.931} & \uline{2.304}
            & \uline{61.139} & \textbf{60.032} \\
        \bottomrule
        \end{tabular}
        }
    \end{minipage}

\end{figure*}

\noindent\textbf{Various Noise-Level Reconstruction.} We evaluate models' noise robustness under Gaussian noise with magnitudes from $1/255$ to $24/255$ as shown in Figure~\ref{fig_noise-level}. Most models suffer severe performance degradation under stronger noise, e.g., DAC drops from 0.943 on clean input to 0.770 at the $24/255$ noise level. In contrast, ROSETok-Base remains relatively stable and still achieves good reconstruction quality under strong noise (24/255), with a 0.920 STOI score.

\subsection{Ablation Study}

\begin{figure*}[t]
  \centering
  \begin{subfigure}[b]{0.245\textwidth}
    \includegraphics[width=\textwidth]{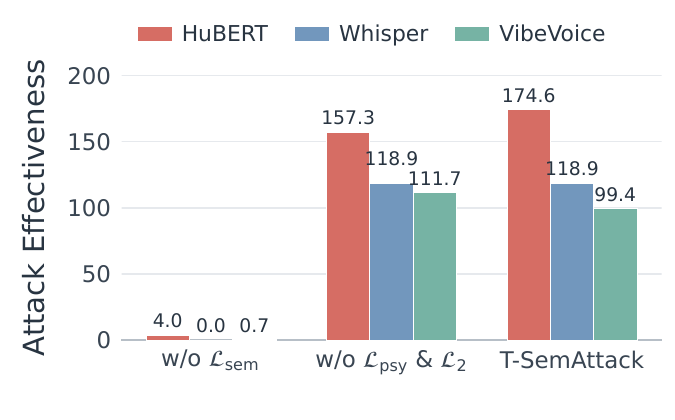}
    \caption{Attack}
    \label{fig:ablation-comp-atk}
  \end{subfigure}
  \hfill
  \begin{subfigure}[b]{0.245\textwidth}
    \includegraphics[width=\textwidth]{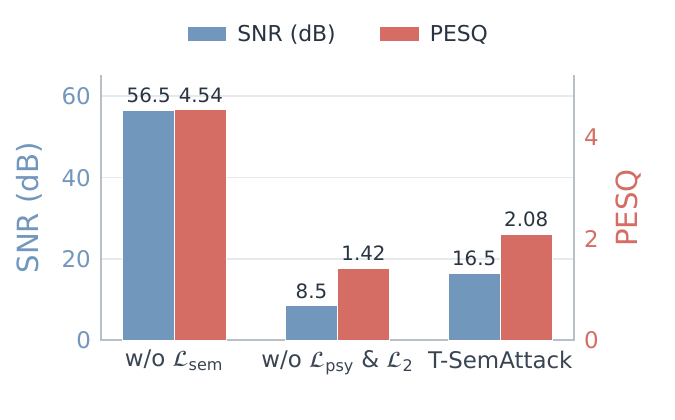}
    \caption{Perception}
    \label{fig:ablation-comp-pcp}
  \end{subfigure}
  \hfill
  \begin{subfigure}[b]{0.245\textwidth}
    \includegraphics[width=\textwidth]{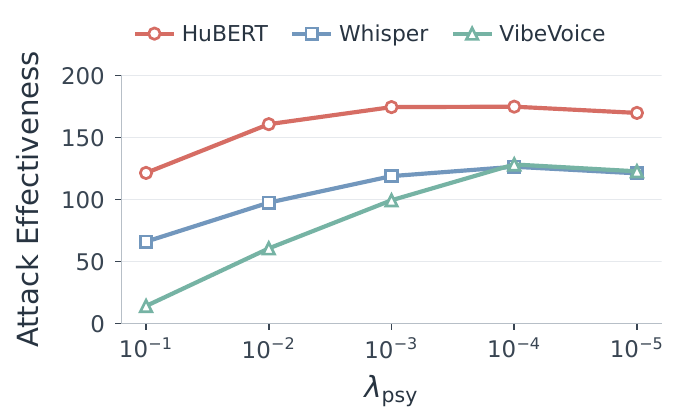}
    \caption{$\lambda_{\mathrm{psy}}$ (Attack)}
    \label{fig:ablation-lam-atk}
  \end{subfigure}
  \hfill
  \begin{subfigure}[b]{0.245\textwidth}
    \includegraphics[width=\textwidth]{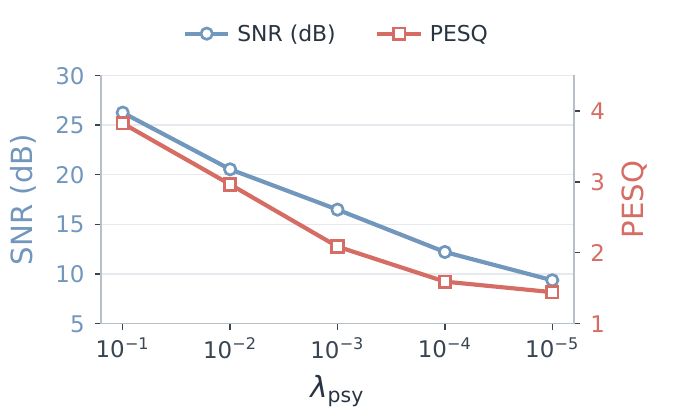}
    \caption{$\lambda_{\mathrm{psy}}$ (Perception)}
    \label{fig:ablation-lam-pcp}
  \end{subfigure}
  \caption{Ablation study for the T-SemAttack. (a)(b) Component Effect. (c)(d) Sensitivity to $\lambda_{\mathrm{psy}}$.}
  \label{fig:ablation}
\end{figure*}

In this section, we conduct ablation experiments. For the attack, we randomly select 30 samples from Seed-TTS (EN). For ROSETok, we train on the LibriSpeech~\cite{librispeech} for 100k steps.

\noindent\textbf{Component Effect of T-SemAttack.} We study the effect of each component in Eq. (\ref{eq:definition}) of T-SemAttack. Results are shown in Figures \ref{fig:ablation-comp-atk} and \ref{fig:ablation-comp-pcp}.
We find that, after removing the semantic attack loss $\mathcal{L}_{\mathrm{sem}}$, the perceptual quality remains high, but the attack is almost ineffective on HuBERT, Whisper, and VibeVoice-ASR, with an attack success rate of only about 1.556\%. After removing the perceptual loss $\mathcal{L}_{\mathrm{psy}}$ and $\mathcal{L}_{2}$, the SNR drops to 8.5, and the PESQ score falls below 2.08. Combining both components achieves a sound balance between perceptual quality and attack effectiveness.

\noindent\textbf{Perception and Attack Balance.} In Eq. (\ref{eq:definition}), $\lambda_{\mathrm{psy}}$ balances attack effectiveness and perceptual quality. We evaluate its sensitivity over the range from $10^{-1}$ to $10^{-5}$. Results in Figure \ref{fig:ablation-lam-atk}\ref{fig:ablation-lam-pcp} show that when $\lambda_{\mathrm{psy}}$ is set to $10^{-3}$ or smaller, the attack effectiveness on the three models shows no significant improvement, while the SNR continues to decrease. Therefore, we set the $\lambda_{\mathrm{psy}}$ to $10^{-3}$.

\noindent\textbf{Component and Semantic Effect of ROSETok.} We verify the effect of ROSETok's components at noise levels of $2/255 \sim 16/255$. 
As shown in Table \ref{tab:ablation_rosetok}, removing noisy training reduces noisy-input STOI from 0.931 to 0.879 and PESQ from 2.304 to 1.578, confirming the importance of noisy-clean training. 
To evaluate semantic capability and consistency, we adapt the ROSETok encoder to the XARES~\cite{xares} benchmark. Higher benchmark scores are better. On Fluent Speech Commands (FSC), a small-scale ASR task, we train on the clean training set and evaluate on its clean and noisy test sets to assess the model's robustness to noise in downstream scenarios. When the noisy signal input is removed, the FSC score drops from 60.032 to 11.363. Removing semantic supervision leads to even more severe semantic degradation on FSC. 
Overall, ROSETok achieves high-fidelity reconstruction of both clean and noisy audio while preserving strong semantic robustness under noisy conditions.

\noindent\textbf{Effect of the Semantic Teacher.} Semantic teacher encoder is used in the training stage of ROSETok. We test the effect of the semantic teacher in ROSETok-Base by replacing Whisper with HuBERT while keeping the remaining training configuration unchanged. Table~\ref{tab:semantic_teacher} reports reconstruction and denoising performance, together with downstream semantic performance on Fluent Speech Commands (FSC). The two teachers yield comparable reconstruction and denoising scores on clean and noisy inputs. However, the downstream FSC score with XARES~\cite{xares} probe decreases from 73.90 with Whisper to 31.48 with HuBERT. These results support the choice of Whisper as the semantic teacher in our setting, despite the similar reconstruction quality obtained with either teacher.

\begin{table}[htbp]
    \centering
    \setlength{\tabcolsep}{8pt}
    \renewcommand{\arraystretch}{1.1}
    \caption{Effect of the semantic teacher on ROSETok-Base reconstruction and downstream semantic performance.}
    \label{tab:semantic_teacher}
    \resizebox{0.7\linewidth}{!}{
    \begin{tabular}{lcccccccc}
        \toprule
        \multirow{2}{*}{\textbf{Teacher}}
        & \multicolumn{2}{c}{\textbf{Clean inputs}}
        & \multicolumn{2}{c}{\textbf{Noisy inputs}}
        & \textbf{FSC (MLP)} \\
        \cmidrule(lr){2-3} \cmidrule(lr){4-5} \cmidrule(lr){6-6}
        & STOI($\uparrow$) & PESQ($\uparrow$) 
        & STOI($\uparrow$) & PESQ($\uparrow$) & FSC($\uparrow$) \\
        \midrule
        Whisper
        & 0.965 & 3.063
        & 0.950 & 2.431
        & \textbf{73.90} \\
        HuBERT
        & 0.965 & 3.088
        & 0.950 & 2.453
        & 31.48 \\
        \bottomrule
    \end{tabular}
    }
\end{table}

\section{Conclusion}
In this work, we explore the robustness of speech tokenizers. Based on the semantic encoder, we propose T-SemAttack, which effectively attacks current mainstream ASR and LALM systems. We further analyze the causes of the attack and, inspired by these findings, propose a robust speech tokenizer that enables high-quality reconstruction from noisy speech. These findings highlight the speech tokenizer as a critical safety bottleneck in modern speech foundation models. We hope our analysis and ROSETok can provide useful insights for developing future speech tokenizers with stronger semantic consistency and robustness.

\section*{Acknowledgments and Disclosures}
We sincerely thank the reviewers for their valuable suggestions. This work is supported by National Natural Science Foundation of China (62076144), Shenzhen Science and Technology Program (JCYJ20220818101014030), and Fundamental Research Funds for the Central Universities under Grant No. 2024ZCJH05. The authors declare no competing interests relevant to this work.

\normalem
\bibliographystyle{unsrt}
\bibliography{references}


\clearpage
\appendix

\section{Discussion and Limitations} \label{sec:discussion}
In this section, we discuss some important aspects.

\textbf{Q1: Can T-SemAttack's perturbation be really imperceptible to humans?}

\textbf{A:} In the setting considered in this paper, what we aim to emphasize is that the noise does not compromise the normal usability of the original audio, that is, its \textbf{harmlessness}, rather than true imperceptibility. T-SemAttack adopts a linear additive strategy and incorporates the masking effect to minimize its impact on the original audio content as much as possible. Subjective experiments in Appendix \ref{sec:human_study} show that T-SemAttack achieves better audio quality and content consistency than baselines, achieving acceptable and normal use of the attacked audio with a MOS-Q of 3.01. Prior research~\cite{safespeech} suggests that MOS scores above 3 indicate noise levels acceptable to human listeners. T-SemAttack's noise level is comparable to that of \textbf{a typical cafe or street environment} where people routinely listen to speech.
In addition, T-SemAttack exhibits strong robustness, which ensures that even if the defender perceives the embedded noise, denoising methods, adversarial defenses, and data transformations cannot effectively remove or neutralize it. Even when adversarial audio is detected, defenders lack effective mitigation strategies. Moreover, because the audio must be retained, e.g., recordings in legal contexts, defenders cannot remove the perturbations by replacing it with a new recording.
{\it \uline{Therefore, T-SemAttack aims to avoid disrupting the normal use of audio, rather than to achieve true imperceptibility}}.

\textbf{Q2: How efficient is T-SemAttack?}

\textbf{A:} First, we focus on an offline scenario, where the real-time requirement is relatively low. In Table \ref{tab:asr_libritts}, we evaluate the RTF of our method and the baselines. Specifically, the RTF is calculated as the total time required by each method to attack the same LibriTTS dataset divided by the total duration of the audio. Details of efficiency computation are provided in Appendix \ref{sec:appendix_cost}.

T-SemAttack achieves an RTF of 21.8, outperforming E2E-VGuard of 26.797. This is because our attack targets the semantic encoder rather than the entire model, which partially improves efficiency. 
It is worth noting that AudioShield is a real-time attack framework, but it requires a very long initialization stage. Specifically, before AudioShield can be used to attack audio, it must first be trained on LibriSpeech for an extended period of time. Therefore, the RTF of AudioShield is 87.045 for warmup-based training and 0.040 for each sample attack after long-time offline training.

Therefore, although T-SemAttack currently achieves better time efficiency than E2E-VGuard, improving its real-time performance remains a direction for future work, e.g., multi-GPU parallelization.

\textbf{Q3: Why does ROSETok preserve semantics at the encoder-level rather than the token-level?}

\textbf{A:} Token-level stability and encoder-level semantic preservation address different failure modes. Token consistency encourages discrete-code invariance, while ROSETok targets the semantic representation drift identified by T-SemAttack. Since token changes do not always imply semantic degradation, we align encoder representations with clean semantic teachers rather than enforcing exact token invariance. Specifically, we consider two reasons.

First, this approach has broader applicability. Some recent codecs~\cite{flexicodec} introduce semantic distillation in the first few layers of RVQ and use the subsequent layers to model acoustic information. Other methods~\cite{stabletoken} directly constrain the stability of quantized tokens under noisy inputs. In contrast, ROSETok directly constrains the semantic representations output by the encoder, encouraging them to remain close to the semantic representations of clean speech under noisy inputs. This not only provides more robust inputs for the subsequent quantizer, but also makes the approach applicable to recent speech tokenizers~\cite{dashengtokenizer} that are based on continuous representations rather than discrete tokens. Accordingly, ROSETok is designed precisely to align the semantic representations of noisy inputs with those of clean audio using a Whisper teacher.

Second, changes in tokens do not necessarily imply changes in semantics. Random Gaussian noise may cause a large proportion of the tokenizer’s output tokens to change. However, if the deep semantic representations remain stable, downstream ASR or LALM models may still recover the correct semantics through their own modeling capabilities. By contrast, the key mechanism of T-SemAttack is not simply to change token IDs, but to drive semantic drift in the encoder representations, making it difficult for downstream models to recover the original content.

Therefore, the goal of ROSETok is not to keep tokens unchanged, but to align the semantic representations of perturbed inputs with those of clean speech. Compared with token-level consistency, encoder-level semantic preservation corresponds more directly to the attack mechanism identified in this paper and, in this work, serves as a more suitable training objective for robust speech tokenizers.

\section{Ethical Considerations and Broader Impacts}\label{sec:ethics}
\noindent\textbf{Dual-Use and Disclaimer of the T-SemAttack.} We recognize that T-SemAttack is a dual-use technique. 
While this work exposes security threats in existing speech tokenizers and encourages researchers to develop more robust speech tokenizers, the same technique could be misused to degrade automatic transcription, evade speech-based moderation or auditing systems, interfere with voice assistants, or reduce the reliability of speech interfaces used in accessibility, documentation, or security-sensitive applications. We therefore treat the proposed attack as a security research artifact rather than a deployable tool.
To mitigate unforeseen impacts, before releasing the source code, we will publish a disclaimer stating that our tool is intended solely for academic research and related purposes. Users must not use this tool for unauthorized or illegal activities, and all actions taken by users are unrelated to the creators and publishers of the tool.

\noindent\textbf{Use of Third-Party Systems.} Our black-box evaluation uses commercial ASR and LALM APIs only to assess transferability. We do not attempt to bypass authentication, rate limits, monitoring, or any access-control mechanism. The evaluation is conducted at a limited scale, uses only public benchmark utterances, and does not involve private user speech. We report aggregate performance metrics and do not use the services to extract model parameters, training data, or proprietary information.

\noindent\textbf{Subjective Experiments.} We obtain informed consent from all participants during recruitment, and all selected participants are over 18 years old. The identities of the participants are anonymized. We do not collect any information beyond what is necessary for the subjective experiments, and the relevant information and responses are used only for academic research. We will not disclose any such sensitive information. All adversarial audio samples are discarded after the subjective experiments are completed, so they will not cause any additional commercial or societal impact.

\textbf{Defensive Intent and Limitations.} We also release and evaluate ROSETok to encourage defensive research on robust speech tokenizers. Nevertheless, ROSETok should not be interpreted as a complete mitigation against all semantic attacks. 
Its robustness depends on the training distribution and threat model, and deployed systems should combine tokenizer-level robustness with additional monitoring, denoising, adversarial evaluation, and system-level safeguards.

\section{Case Study: T-SemAttack in Text-To-Speech for Protection}\label{sec:appendix_case_study}

To evaluate the potential application of T-SemAttack in speech protection scenarios, we conduct a case study on text-to-speech (TTS) systems. Specifically, we consider a realistic threat in which an attacker can obtain a user’s speech samples and utilize advanced TTS systems for voice cloning~\cite{e2e-vguard}. If the original speech is slightly perturbed before being released or uploaded in a form that remains acceptable to humans, downstream speech generation systems would struggle to accurately extract speaker characteristics or semantic information. Motivated by this scenario, we evaluate whether T-SemAttack can serve as a speech protection against the unauthorized use of TTS systems.

We select three representative TTS systems for evaluation, i.e., CosyVoice~\cite{cosyvoice}, CosyVoice2~\cite{cosyvoice2}, and Step-Audio-TTS~\cite{step-audio}. We evaluate the generated speech from two perspectives: WER and SIM. A higher WER indicates more severe semantic deviation in the generated content, while a lower SIM indicates that the system has greater difficulty preserving the voice characteristics of the original speaker. We conduct experiments on the Seed-TTS (EN) dataset, using the ground-truth transcription of each audio sample as the input reference text.

\begin{table}[h]
    \centering
    \caption{Case study of T-SemAttack in the TTS protection scenario. The higher WER(\%) represents more severe semantic errors in the generated speech. The lower the SIM, the lower the similarity between the generated speech and the original speaker. }
    \label{tab:tts_protection_case}
    \begin{tabular}{lcccccc}
    \toprule
    \multirow{2}{*}{Method} 
    & \multicolumn{2}{c}{CosyVoice~\cite{cosyvoice}} 
    & \multicolumn{2}{c}{CosyVoice2~\cite{cosyvoice2}} 
    & \multicolumn{2}{c}{Step-Audio-TTS~\cite{step-audio}} \\
    \cmidrule(lr){2-3} \cmidrule(lr){4-5} \cmidrule(lr){6-7}
    & WER($\uparrow$) & SIM($\downarrow$)
    & WER($\uparrow$) & SIM($\downarrow$)
    & WER($\uparrow$) & SIM($\downarrow$) \\
    \midrule
    Clean        & 4.583  & 0.599 & 4.484  & 0.836 & 3.734   & 0.772 \\
    \textbf{T-SemAttack}  
        & \textbf{52.476} & \textbf{0.212} 
        & \textbf{33.427} & \textbf{0.330} & \textbf{122.746} & \textbf{0.299} \\
    \bottomrule
    \end{tabular}
\end{table}

Table~\ref{tab:tts_protection_case} presents the results of this case study. With clean input, all three TTS systems preserve the semantic content well, achieving WERs of only 4.583, 4.484, and 3.734, respectively. These observations suggest that these systems can effectively use the speaker features contained in the input speech for speech generation. In contrast, when the input speech is processed by T-SemAttack, the generation quality of all TTS systems degrades substantially. Step-Audio-TTS is affected most severely, with its WER increasing sharply from 3.734 to 122.746. On average, the WER increases from 4.267 under the clean condition to 69.550, while the speaker similarity decreases from 0.736 to 0.280. These results show that T-SemAttack can disrupt the modeling of speech content in TTS.

This result suggests that semantic-level perturbations can also affect the conditional modeling process in TTS systems. When a TTS system relies on input speech to extract semantic or speaker representations, the semantic representation disruption and tokenizer-level representation perturbation caused by T-SemAttack can affect the generation stage, resulting in content errors and reduced speaker similarity in the generated speech. Therefore, this case study provides another perspective supporting the critical role of speech tokenizers in modern speech systems.

\section{Human Study}\label{sec:human_study}
To evaluate the human perception of the perturbation, we conduct a subjective human listening study to assess the perceptual quality of the attacked audio generated by T-SemAttack and the baselines. This section details the recruitment process, design of the questionnaire, and results of evaluations.

\textbf{Recruitment Process.} This subjective survey has been approved by the Human Ethics Research Committee at the first author's institution. We create the questionnaire through the Credamo platform and recruit 20 volunteers to participate in the survey, which is comparable to similar studies, e.g., AntiFake (24 participants) and E2E-VGuard (36 participants). All volunteers are over 18 years old and are English-educated. Their average response time is $\sim$330 seconds, providing reliable results.

\textbf{Questionnaire Setup.} We randomly select four speakers (two male and two female) from the Seed-TTS-EN dataset, with one utterance per speaker. Each utterance is processed by baselines, i.e., E2E-VGuard and AudioShield , as well as our T-SemAttack, resulting in 12 attacked audio samples and 4 original samples. For each attacked sample, participants are asked to compare it with the corresponding original audio and provide ratings from three perspectives: (1) \textbf{Audio Quality}, which measures the noise level and overall perceptual quality of the attacked audio; (2) \textbf{Content Consistency}, which evaluates whether the linguistic content of the attacked audio remains aligned with the original; and (3) \textbf{Timbre Similarity}, which assesses whether the speaker identity in the attacked audio is preserved relative to the original. 

All three aspects are scored on a 0$\sim$5 point Mean Opinion Score (MOS) scale, where higher scores indicate better perceptual quality. To improve the confidence of the subjective evaluation and reduce bias, we report the MOS values together with their 95\% confidence intervals~\cite{safespeech}.

\begin{table}[h]
    \centering
    \caption{Subjective evaluation from audio quality, content intelligibility, and speaker similarity aspects with 95\% confidence intervals.}
    \label{tab:human_study}
    \begin{tabular}{lccc}
    \toprule
    \textbf{Method} & \textbf{MOS-Q}($\uparrow$) & \textbf{MOS-C}($\uparrow$) & \textbf{MOS-S}($\uparrow$) \\
    \midrule
    E2E-VGuard                  & 2.67$\pm$0.21 & 4.29$\pm$0.17 & \textbf{3.98$\pm$0.20} \\
    AudioShield  & 2.63$\pm$0.30 & 3.32$\pm$0.31 & 1.66$\pm$0.33 \\
    \textbf{T-SemAttack}        & \textbf{3.01$\pm$0.28} & \textbf{4.32$\pm$0.19} & 3.87$\pm$0.25 \\
    \bottomrule
    \end{tabular}
\end{table}

\textbf{Results and Analysis.} The detailed MOS results and analysis are provided in Table~\ref{tab:human_study}. We evaluate three dimensions: overall quality (MOS-Q), content intelligibility (MOS-C), and speaker similarity (MOS-S). Across all three dimensions, our method, T-SemAttack, consistently achieves the highest or comparable MOS values among all methods. In particular, AudioShield introduces a clearly perceivable perturbation that almost prevents listeners from recognizing the linguistic content of the original audio, as reflected by its low MOS-Q (2.63), MOS-C (3.32), and MOS-S (1.66). We attribute this to the fact that AudioShield is trained on LibriSpeech and has fitted a substantial amount of female voice characteristics, and that modifying the latent representation cannot guarantee the fidelity of the reconstructed waveform. E2E-VGuard better preserves content intelligibility with a MOS-C of 4.29 and a MOS-S of 3.98 for speaker similarity, but still introduces audible artifacts that lower its overall quality with a MOS-Q of 2.67. In contrast, T-SemAttack adopts a linear additive strategy combined with the psychoacoustic masking effect, which minimizes its impact on the original audio content. It achieves the highest MOS-Q (3.01), \textit{which surpasses the 3.0 threshold generally regarded as acceptable perceptual distortion}~\cite{safespeech}, and maintains strong MOS-C (4.32) and MOS-S (3.87). 
These subjective experiments show that T-SemAttack does not compromise the linguistic content or speaker timbre, thereby preserving the normal usability of the audio.

\begin{figure}[h]
    \centering
    \includegraphics[width=0.8\textwidth]{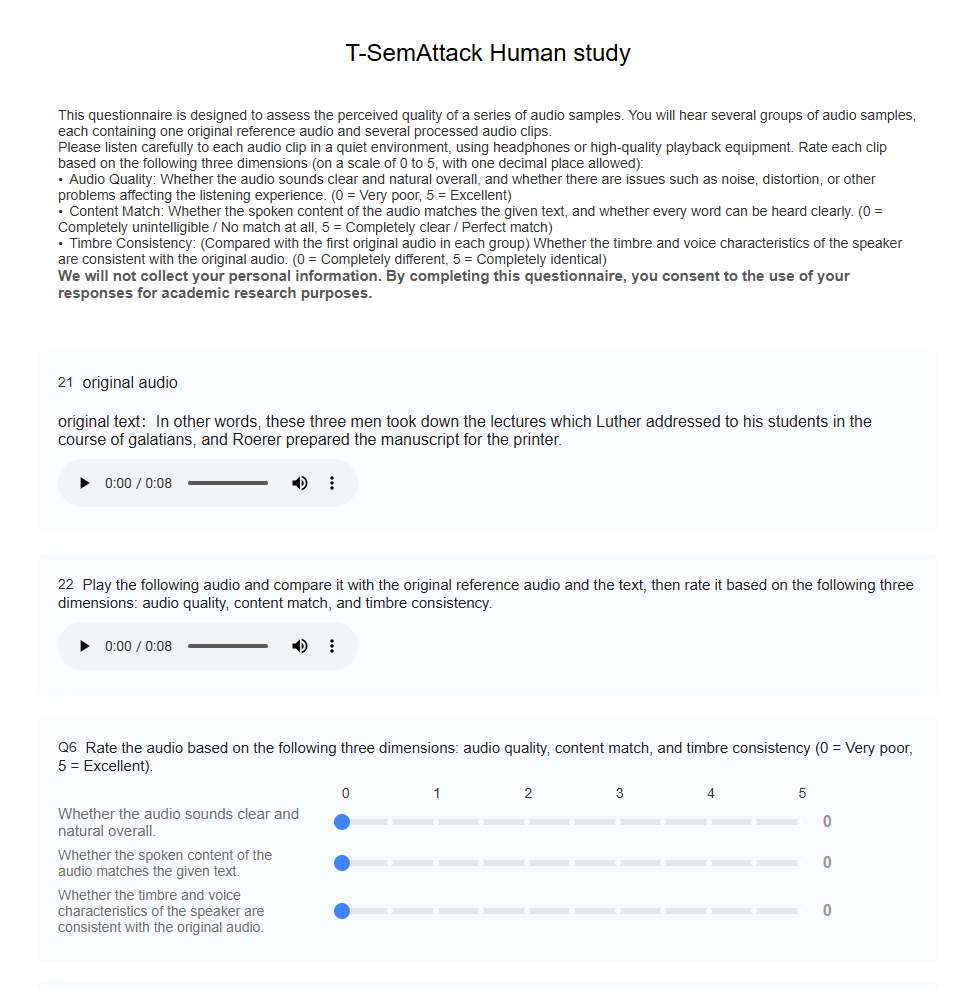}
    \caption{The screenshot of the questionnaire for subjective experiments.}
    \label{fig:questionnaire}
\end{figure}

The screenshot and full text of the questionnaire are shown in Figure \ref{fig:questionnaire}.

\section{Additional Details}\label{sec:details}

\subsection{Attack Details}
Table \ref{tab:attack_details} summarizes the main hyperparameters used throughout T-SemAttack.
\begin{table}[h]
    \centering
    \caption{Hyperparameters of T-SemAttack.}
    \label{tab:attack_details}
    \begin{tabular}{ll} 
    \toprule
    Hyperparameter & Value \\
    \midrule
        lr             & $1\mathrm{e}{-3}$ \\
        lr\_min        & $1\mathrm{e}{-6}$ \\
        max\_steps     & 500 \\
        $\epsilon$     & $8/255$ \\
    \bottomrule
    \end{tabular}
\end{table}
``lr'' refers to the initial learning rate of the SGD optimizer, used to control the update step size of the adversarial noise during backpropagation.
``lr\_min'' refers to the minimal learning rate. ``max\_steps'' denotes the total optimization steps for the T-SemAttack. $\epsilon$ represents the perturbation amplitude.

\subsection{Adversarial Defense Details}\label{appendix:details_adv_defense}
The following are the details of adversarial defense and audio transformations for Table \ref{tab:defense_transformation} we use against semantic attacks:
\begin{itemize}[leftmargin=*]
    \item \textbf{Resample.} Following standard adversarial defense practice~\cite{waveguard}, we randomly downsample the input waveform to one of {8000, 10000, 12000} Hz and then upsample it back to the original 24 kHz. This destroys high-frequency components where adversarial perturbations are often concentrated.
    \item \textbf{Quantize and Dequantize (Q-D).} The waveform is linearly mapped from the [-1, 1] range to a b-bit integer range and then mapped back to its original scale. This bit-depth reduction acts as a coarse projection that disrupts small adversarial perturbations in the waveform amplitude.
    \item \textbf{Mel-Inverse Reconstruction (Mel)}. The waveform is first transformed into a mel-spectrogram and then reconstructed back to the time domain using a Griffin–Lim–based inversion. This process is lossy and discards the original phase information, acting as a strong non-linear filter that removes perturbations not aligned with the mel-scale spectral structure. The configuration uses an FFT size of 2048, a hop length of 512, 128 mel bands, and a sampling rate of 24 kHz.
    \item \textbf{AudioPure}~\cite{audiopure}. AudioPure is a defense method against adversarial audio attacks. It uses a diffusion model to purify perturbed audio by adding noise and then denoising it through the reverse diffusion process, reducing adversarial perturbations and making the audio closer to clean speech, thereby improving the robustness of speech recognition or audio classification systems under attack.
    \item \textbf{Change speed (Speed).} We apply time stretching with a randomly selected rate from the set {0.8, 0.9, 1.0, 1.1, 1.2}. This temporally realigns acoustic events without altering pitch.
    \item \textbf{MP3 Compression (MP3).} We apply MP3 compression with bitrates uniformly sampled between 8 and 64 kbps while preserving the original waveform length. The compressed signal is either truncated or zero-padded as needed to maintain alignment.
    \item \textbf{Pitch Shift (PS).} We shift the pitch by a random number of semitones uniformly sampled from -4 to +4. This modifies the pitch contour while preserving the duration of the audio.
    \item \textbf{Band-Pass Filter (BPF).} We apply a band-pass filter with a center frequency uniformly sampled between 100 and 6000 Hz. This restricts the spectral content of the input, retaining only a specific frequency band.
\end{itemize}

The adversarial defense methods follow WaveGuard~\cite{waveguard}. Audio transformation techniques are implemented using the audiomentations library.

\subsection{ROSETok and Training Details} \label{sec:appendix_details_training}

\noindent\textbf{Details of the ROSETok Architecture.}
The ROSETok framework comprises several core components: an encoder compresses the input audio waveform into a 1280-dimensional latent representation, which is then aligned with semantic embeddings from a Whisper Teacher model. A 9-codebook Residual Vector Quantizer subsequently discretizes this latent representation into compact codes, and a decoder reconstructs the waveform from these codes.

The DAC encoder and decoder are constructed from symmetrically arranged downsampling and upsampling modules. The encoder contains four EncoderBlocks that progressively downsample the signal with striding factors of 2, 4, 5, and 8, while expanding the channel dimension from 64 to 1024 and finally projecting to a 1280-dimensional latent space. Each EncoderBlock sequentially connects three residual units with dilation rates of 1, 3, and 9, followed by a Snake activation function. The decoder mirrors this architecture exactly, starting with 1536 channels and employing four DecoderBlocks with upsampling factors of 8, 5, 4, and 2 to restore the monaural audio waveform.

The quantizer adopts residual vector quantization with nine codebook layers, where each layer maintains 1024 entries of 8-dimensional vectors. The quantization dropout rate is set to zero to ensure training stability. Through successive residual refinement across codebook layers, a compact 8-dimensional discrete representation approximates the original 1280-dimensional continuous latent vector, achieving a compression ratio of approximately 17.8.

During the training stage, two additional components are introduced, namely the semantic encoder and the discriminator. During the training process, the semantic encoder employs a frozen Whisper large-v3 encoder, which first extracts Mel spectrogram features via short-time Fourier transform and then produces 1280-dimensional semantic embeddings through its Transformer backbone. This embedding dimensionality aligns precisely with the output dimension of the DAC encoder. Additionally, the discriminator is introduced to enhance the perceptual quality of the reconstructed audio by comparing it against the original audio, where adversarial loss and feature matching loss are computed to guide the training.

\noindent\textbf{Details of Training Objectives.}
The details of each loss during training are as follows:

$\mathcal{L}_{\mathrm{distill}}$ is the semantic distillation loss, which forces the audio encoder to learn semantically rich representations by maximizing the cosine similarity between the latent representations output by the encoder and the semantic embeddings extracted by the frozen Whisper teacher model~\cite{whisper}.
The semantic distillation loss $\mathcal{L}_{\mathrm{distill}}$ is computed as $\mathcal{L}_{\mathrm{distill}} = 1 - \text{CosSim}(z, z_{\mathrm{sem}})$, where $z$ denotes the encoded representation and $z_{\mathrm{sem}}$ is the representation from the semantic encoder.

$\mathcal{L}_{\mathrm{recon}}$ is the reconstruction loss, which ensures the fidelity of the codec process by minimizing the difference between the audio reconstructed by the decoder and the original audio.

$\mathcal{L}_{\mathrm{cb}}$ is the codebook loss, which optimizes the codebook vectors of the residual vector quantizer to better fit the feature distribution output by the encoder.

$\mathcal{L}_{\mathrm{commit}}$ is the commitment loss, which constrains the encoder output to stay close to the nearest codebook vector, preventing large fluctuations between different codebooks.

$\mathcal{L}_{\mathrm{adv}}$ is the adversarial loss, which improves the perceptual quality of the generated audio by training a discriminator to distinguish between real and reconstructed audio.

$\mathcal{L}_{\mathrm{fm}}$ is the feature matching loss, which stabilizes the adversarial training process and further enhances the quality of the reconstructed audio by minimizing the difference between features extracted from the original and reconstructed audio at intermediate layers of the discriminator.

We adopt $\mathcal{L}_{\mathrm{recon}}$, $\mathcal{L}_{\mathrm{cb}}$, and $\mathcal{L}_{\mathrm{commit}}$ from \cite{dac}, and $\mathcal{L}_{\mathrm{adv}}$ and $\mathcal{L}_{\mathrm{fm}}$ from \cite{llasa}.

\noindent\textbf{Details of Training Datasets.} We train the ROSETok model on 96.7k hours of open-source English and Chinese data. 
All open-source datasets used in this work are listed in Table \ref{tab:train_details}. LibriSpeech is used for ablation and the stage-two fine-tuning.

\begin{table}[h]
    \centering
    \caption{Detailed information of training datasets.}
    \label{tab:train_details}
    \begin{tabular}{l r r r r r}
    \toprule
    Dataset & Subset & Duration (\# h) & Task & Language & Usage \\
    \midrule
    Librispeech
        & train \& val        & 960       & ASR        & English & Ablation \& ROSETok-AAF \\
    Emilia
        &  EN & 46,800      & ASR    & English & ROSETok-Base \\
    Emilia
        &  ZH & 49,900      & ASR    & Chinese & ROSETok-Base \\
    \bottomrule
    \end{tabular}
\end{table}
    
\noindent\textbf{Details of Hyperparameters.} Table \ref{tab:train_hyperparams} summarizes the main hyperparameters used throughout ROSETok training.
``lr'' refers to the initial learning rate of the AdamW optimizer, controlling the step size of parameter updates during training.
``lr\_scheduler'' refers to the learning rate scheduling strategy, where CosineLR is used to smoothly anneal the learning rate following a cosine curve.
``batch\_size'' refers to the global number of audio samples per training step. 
``noise\_level'' refers to the noise amplitude during noisy training.
``warmup\_steps'' refers to the initial $1000$ steps during which the learning rate is linearly warmed up to stabilize early-stage training.
``total\_steps'' refers to the total number of optimization steps $200{,}000$ for training the robust audio tokenizer.

\begin{table}[h]
    \centering
    \caption{Training Hyperparameters of the ROSETok.}
    \label{tab:train_hyperparams}
    \begin{tabular}{ll} 
    \toprule
    \textbf{Hyperparameter}  & \textbf{Value} \\
    \midrule
    \multicolumn{2}{l}{\textit{ROSETok-Base}} \\
    optimizer & AdamW \\
    lr              & $1.0 \times 10^{-4}$ \\
    lr\_scheduler   & CosineLR \\
    min\_lr         & $1.0 \times 10^{-5}$ \\
    batch\_size     & 128 \\
    noise\_level    & $4 / 255 \sim 24 / 255$ \\
    warmup\_steps   & 1000 \\
    total\_steps    & 200,000 \\
    \midrule

    \multicolumn{2}{l}{\textit{ROSETok-AAF}} \\
    optimizer & AdamW \\
    lr              & $5.0 \times 10^{-6}$ \\
    lr\_scheduler   & CosineLR \\
    min\_lr         & $5.0 \times 10^{-7}$ \\
    batch\_size     & 64 \\
    warmup\_steps   & 100 \\
    total\_steps    & 2,000 \\
    
    \bottomrule
    \end{tabular}
\end{table}

\noindent\textbf{Details of the XARES Benchmark.} 
XARES~\cite{xares} evaluates audio encoders through downstream probing tasks. We freeze the encoder, extract frame-level speech representations, and feed them into lightweight task-specific MLP heads for classification. Following the XARES protocol, raw task metrics are normalized to $[0,1]$, where higher scores indicate stronger semantic capability. For readability, we report the normalized scores in percentage form.

\subsection{Baseline and Model Details}
The baseline attack methods considered in our study are as follows:
(1) \textbf{Gaussian Noise.} As the most fundamental comparison baseline, this method introduces perturbations by adding random Gaussian noise to the original audio signal. To ensure that the signal-to-noise ratio of the processed audio is comparable to that of our proposed method, facilitating a fair comparison, we set the noise standard deviation to 0.01. (2) \textbf{E2E-VGuard}~\cite{e2e-vguard}
is an end-to-end adversarial defense scheme designed for speech models. In our experiments, we strictly follow the default parameter configurations from the original paper to ensure the reliability of reproduction results and the objectivity of the comparison. (3) \textbf{AudioShield}~\cite{audioshield}
attacks audio encoders by applying perturbations in the latent feature space, where a threshold parameter $\tau$ controls the trade-off between perturbation strength and perception. We select the $\tau$ = 0.1 to achieve a comparable perceptual results with T-SemAttack.

To comprehensively evaluate the effectiveness of our semantic attack, we test the adversarial audio against a diverse set of Automatic Speech Recognition (ASR) models spanning both open-source and closed-source systems. The open-source models we use are as follows:

(1) \textbf{HuBERT-Large}
~\cite{hubert} is a self-supervised speech representation model fine-tuned on 960 hours of LibriSpeech data, known for its strong performance on English transcription tasks through masked prediction of hidden units. 
(2) \textbf{Whisper-Large}
~\cite{whisper} is OpenAI's multilingual, multitask ASR model trained on 680k+ hours of weakly supervised audio data, offering robust performance across languages and acoustic conditions. HuBERT-Large and Whisper-Large are evaluated as ASR systems in Table \ref{tab:asr_libritts}, while their encoders are also used as surrogate branches during attack optimization. 
(3) \textbf{Qwen3-ASR-0.6B}
~\cite{qwen3-asr} is Alibaba's lightweight ASR model built upon the Qwen3 foundation, providing efficient and accurate multilingual speech recognition with a compact 0.6B parameter footprint. 
(4) \textbf{VibeVoice-ASR}
~\cite{vibevoice-asr} is a recent ASR system designed for natural conversational speech, emphasizing robustness to spontaneous and expressive vocal styles. 
(5) \textbf{Fun-ASR-Nano}
~\cite{fun-asr} is a compact variant of the FunASR family developed by Alibaba DAMO Academy, optimized for low-latency inference while maintaining competitive accuracy on Mandarin and multilingual benchmarks. 
(6) \textbf{FireRedASR}
~\cite{fireredasr} is an attention-based encoder-decoder ASR model from Xiaohongshu, achieving SOTA results on Mandarin benchmarks with strong generalization to diverse acoustic scenarios. 

For closed-source ASR systems, we evaluate against three widely-used commercial APIs: iFlytek, Doubao, and GPT-4o-mini-transcribe, which allows us to assess the transferability of our attack to black-box production systems. 
(1) \textbf{iFlytek} is a leading Chinese commercial ASR service renowned for its high accuracy on Mandarin and various dialects in industrial deployments. 
(2) \textbf{Doubao-ASR} is ByteDance's commercial speech recognition API, offering robust transcription capabilities across multiple languages and noisy real-world conditions. 
(3) \textbf{GPT-4o-mini-transcribe} is OpenAI's lightweight transcription endpoint derived from the GPT-4o family, providing high-quality multilingual transcription through a unified multimodal architecture.

In addition to dedicated ASR systems, we further evaluate our attack against LALMs, which integrate audio understanding with large language model capabilities. 
(1) \textbf{Qwen2-Audio}
~\cite{qwen2-audio} is Alibaba's instruction-tuned audio-language model that supports both voice chat and audio analysis, capable of processing diverse audio inputs including speech, sounds, and music. 
(2) \textbf{Qwen3-Omni}
~\cite{qwen3-omni} is a flagship omni-modal model with a Mixture-of-Experts architecture (30B total / 3B active parameters), unifying text, image, audio, and video understanding with strong instruction-following capabilities. 
(3) \textbf{Audio-Flamingo-3}
~\cite{af3} is NVIDIA's advanced audio-language model that excels at understanding complex audio content, including speech, music, and ambient sounds, with strong reasoning and dialogue capabilities. 
The closed-source category includes \textbf{Gemini-3-Flash-Preview}, Google's latest multimodal model from the Gemini family, offering fast inference and strong cross-modal understanding across audio, vision, and text. 

To ensure these models perform pure transcription rather than higher-level reasoning or summarization, we adopt a unified system prompt of  ``\textit{Transcribe the audio exactly as recorded. Output the transcription only. Do not add explanations, summaries, or comments.}'' This setup enables a fair comparison between LALMs and conventional ASR models under the same transcription objective.

To evaluate the effectiveness of our noise-robust tokenizer, we compare it against a diverse set of representative baseline tokenizers.
(1) \textbf{SNAC}
~\cite{snac} is a multi-scale neural audio codec that employs hierarchical RVQ at different temporal resolutions to efficiently compress audio signals. 
(2) \textbf{XY-Tokenizer}
~\cite{xy-tokenizer} is a speech tokenizer designed to jointly model semantic and acoustic information, enabling unified representation for speech understanding and generation. 
(3) \textbf{XCodec-2}
~\cite{llasa} integrates semantic features from self-supervised models with acoustic reconstruction objectives, producing discrete tokens that balance linguistic content and audio fidelity. 
(4) \textbf{DAC}
~\cite{dac} is a high-fidelity general-purpose neural audio codec that leverages improved RVQ techniques and adversarial training to achieve strong reconstruction quality across diverse audio domains. 
(5) \textbf{MOSS-Audio-Tokenizer}
~\cite{moss} is a neural audio codec that compresses raw audio into discrete tokens, enabling efficient audio generation and understanding within the MOSS multimodal framework. 
(6) \textbf{StableToken}
~\cite{stabletoken} is a tokenizer that focuses on producing stable and consistent discrete representations across varying input conditions, aiming to mitigate token instability in downstream tasks. 
Together, these baselines cover a broad spectrum of design philosophies, ranging from purely acoustic codecs to semantically-enriched and stability-oriented tokenizers, providing a comprehensive baseline system for assessing noise robustness.

\subsection{Compute Reporting and Efficiency Details}\label{sec:appendix_cost}
\noindent\textbf{Details of Attacks' Efficiency Evaluation.}
In Table \ref{tab:asr_libritts}, the RTF is computed as the ratio of the per-sample perturbation generation time to the total duration of the audio in Eq. (\ref{eq:rtf_attacks}).
\begin{equation}
    \text{RTF}_{\text{Attacks}} = \frac{\sum_{i=1}^{N} T_{\text{attack}}^{(i)}}{\sum_{i=1}^{N} T_{\text{audio}}^{(i)}},
    \label{eq:rtf_attacks}
\end{equation}
where $T_{\text{attack}}^{(i)}$ is the time taken by each attack method to process the i-th audio sample, $T_{\text{audio}}^{(i)}$ is the duration of the i-th audio sample, and N is the total number of test samples.

AudioShield constitutes a special case because it pre-trains a universal adversarial perturbation (UAP) on a training set, then directly applies it to all test inputs without per-audio optimization. Since the training stage is essential for obtaining a usable perturbation, we amortize its cost across test samples to ensure a fair comparison in Eq. (\ref{eq:rtf_audioshield}).
\begin{equation}
    \text{RTF}_{\text{AudioShield}} = \frac{T_{\text{train}} + \sum_{i=1}^{N} T_{\text{apply}}^{(i)}}{\sum_{i=1}^{N} T_{\text{audio}}^{(i)}},
    \label{eq:rtf_audioshield}
\end{equation}
where $T_{\text{train}}$ is the total perturbation training time, $T_{\text{apply}}^{(i)}$ is the time to apply the pre-trained perturbation to the i-th audio sample. This amortizes the training overhead across test samples, enabling a fair comparison with other baselines under a unified framework. When $N$ is sufficiently large, the amortized $T_{\text{train}}$ term becomes negligible, and the RTF of AudioShield gradually approaches its theoretical lower bound.
The RTF measurements of attacks are conducted on a device equipped with an NVIDIA 4090 GPU and Intel(R) Xeon(R) Platinum 8488C CPU, 512GB RAM.

\noindent\textbf{Details of Tokenizers' Efficiency Evaluation. }
In Table \ref{tab:tokenizer_robustness}, the Real-Time Factor (RTF) is computed as the ratio of the total audio reconstruction time to the total duration of the audio in Eq. (\ref{eq:rtf_recon}).
\begin{equation}
    \text{RTF}_{\text{Tokenizers}} = \frac{\sum_{i=1}^{N} T_{\text{reconstruct}}^{(i)}}{\sum_{i=1}^{N} T_{\text{audio}}^{(i)}},
    \label{eq:rtf_recon}
\end{equation}
The reconstruction time $T_{\text{reconstruct}}^{(i)}$ measures only the tokenizer's core reconstruction process, namely waveform encoding and decoding. It excludes the time required to load audio files from disk or save reconstructed audio back to disk. This ensures that the reported RTF reflects the robust tokenizer's intrinsic computational efficiency, rather than I/O overhead unrelated to model performance. The RTF measurements of tokenizers are conducted on a device equipped with an NVIDIA 4090 GPU, Intel(R) Xeon(R) Platinum 8488C CPU, and 512 GB RAM.

\noindent\textbf{Compute Reporting.} We conduct our experiments using GPU-equipped computing resources. Specifically, we run T-SemAttack and the evaluation on a single NVIDIA RTX 4090 GPU. For training ROSETok, we use 8 NVIDIA H100 GPUs for approximately 20 hours.

\section{Visualization and Analyses of the T-SemAttack}
\subsection{Perturbation Amplification}\label{sec:appendix_amplification}

To analyze the mechanism of T-SemAttack in the speech tokenizer, we compute the representation differences between clean audio $x$ and adversarial audio $x_{\mathrm{adv}}$ at different internal stages of the tokenizer. We use the S3 Tokenizer as an example, and its audio processing pipeline is: $\text{Waveform} \rightarrow \text{Mel-spectrogram} \rightarrow \text{Encoder} \rightarrow \text{VQ Tokens}$.

For continuous representation stages, we use relative $L_2$ distortion, referred to as Relative Distortion (RD), to measure the perturbation magnitude:
\[
\mathrm{RD}_s =
\frac{\left\| z_s(x_{\mathrm{adv}})-z_s(x) \right\|_2}
{\left\| z_s(x) \right\|_2},
\]
where \(z_s(\cdot) \in \mathbb{R}^{T_s \times d_s}\) denotes the representation at stage \(s\),
\(T_s\) is the sequence length of the representation at that stage, and \(d_s\) is the feature
dimension. Note that \(T_s\) is stage-dependent and may differ from the waveform length \(T\).
We also use cosine distance (CD) to measure changes in the representation direction:
\[
\mathrm{CD}_s=1 -\frac{1}{T_s}\sum_{t=1}^{T_s}\cos\left(z_{s,t}(x), z_{s,t}(x_{\mathrm{adv}})\right).
\]
The amplification factor at each stage is defined as the ratio between the average perturbation at that stage and the average perturbation at the waveform stage:
\[
\mathrm{Amp}_{\mathrm{RD}}(s)=
\frac{\frac{1}{N}\sum_{i=1}^{N}\mathrm{RD}^{(i)}_s}
{\frac{1}{N}\sum_{i=1}^{N}\mathrm{RD}^{(i)}_{\mathrm{wav}}}.
\]
For the final discrete VQ tokens, we compute the token error rate (TER):
\[
\mathrm{TER} =
1 - \frac{\sum_{n=1}^{N}\mathbf{1}[c_n(x)=c_n(x_{\mathrm{adv}})]}{N},
\]
where $c_n(\cdot)$ denotes the $n$-th discrete token and $N$ denotes the number of tokens.

Table~\ref{tab:s3_amplification} reports the perturbation amplification results on the S3 Tokenizer. The average RD at the waveform stage is only 0.154, but it rapidly increases to 0.771 at the Mel-spectrogram stage, corresponding to a $5.01\times$ amplification. After the encoder, the average RD further increases to 1.520, which is $9.87\times$ larger than that at the waveform stage. Meanwhile, the average CD of the encoder output reaches 1.248, indicating a severe directional shift in the high-level semantic representations of the adversarial samples compared with the clean samples.

\begin{table}[t]
    \centering
    \caption{Perturbation amplification results in the S3 Tokenizer.}
    \label{tab:s3_amplification}
    \resizebox{\linewidth}{!}{
    \begin{tabular}{lcccccc}
    \toprule
    Stage & RD Mean & RD Std & RD Median & CD Mean & CD Median & Amp(RD) \\
    \midrule
    Waveform        & 0.154 & 0.023 & 0.153 & 0.322 & 0.326 & $1.00\times$ \\
    Mel Spectrogram & 0.771 & 0.070 & 0.772 & 0.357 & 0.347 & $5.01\times$ \\
    Encoder Output  & 1.520 & 0.055 & 1.530 & 1.248 & 1.250 & $9.87\times$ \\
    \bottomrule
    \end{tabular}
    }
\end{table}

Finally, at the VQ token level, T-SemAttack achieves an average TER of 99.7\%. This shows that perturbation drift in the continuous representation space is ultimately quantized into an almost entirely different discrete token sequence. Therefore, T-SemAttack does not merely introduce local waveform noise. Instead, it creates a progressive amplification chain inside the S3 Tokenizer, from the acoustic frontend to the semantic encoder, ultimately causing tokenizer-level semantic collapse.

\subsection{LALM Attention}\label{sec:lalm_attn}
To further analyze the impact of T-SemAttack on the cross-modal modeling process of LALMs, we take Qwen2-Audio as an example and conduct a layer-wise analysis of Transformer attention under clean and adversarial audio inputs. The analysis consists of a single-sample attention-map visualization, which intuitively shows changes in the attention distribution before and after the attack.

\noindent\textbf{Single-Sample Attention Visualization.}
To illustrate attention disruption, we select one clean-adversarial pair and visualize the attention maps of several representative layers. For each layer, we average over all attention heads and then plot the resulting two-dimensional attention matrix. In the figure, the x-axis denotes key positions, and the y-axis denotes query positions. Since the LLM component of Qwen2-Audio uses causal attention, the future-token region in the upper-right corner is masked, appearing as a black triangular area.

\begin{figure*}[t]
    \centering
    \includegraphics[width=\textwidth]{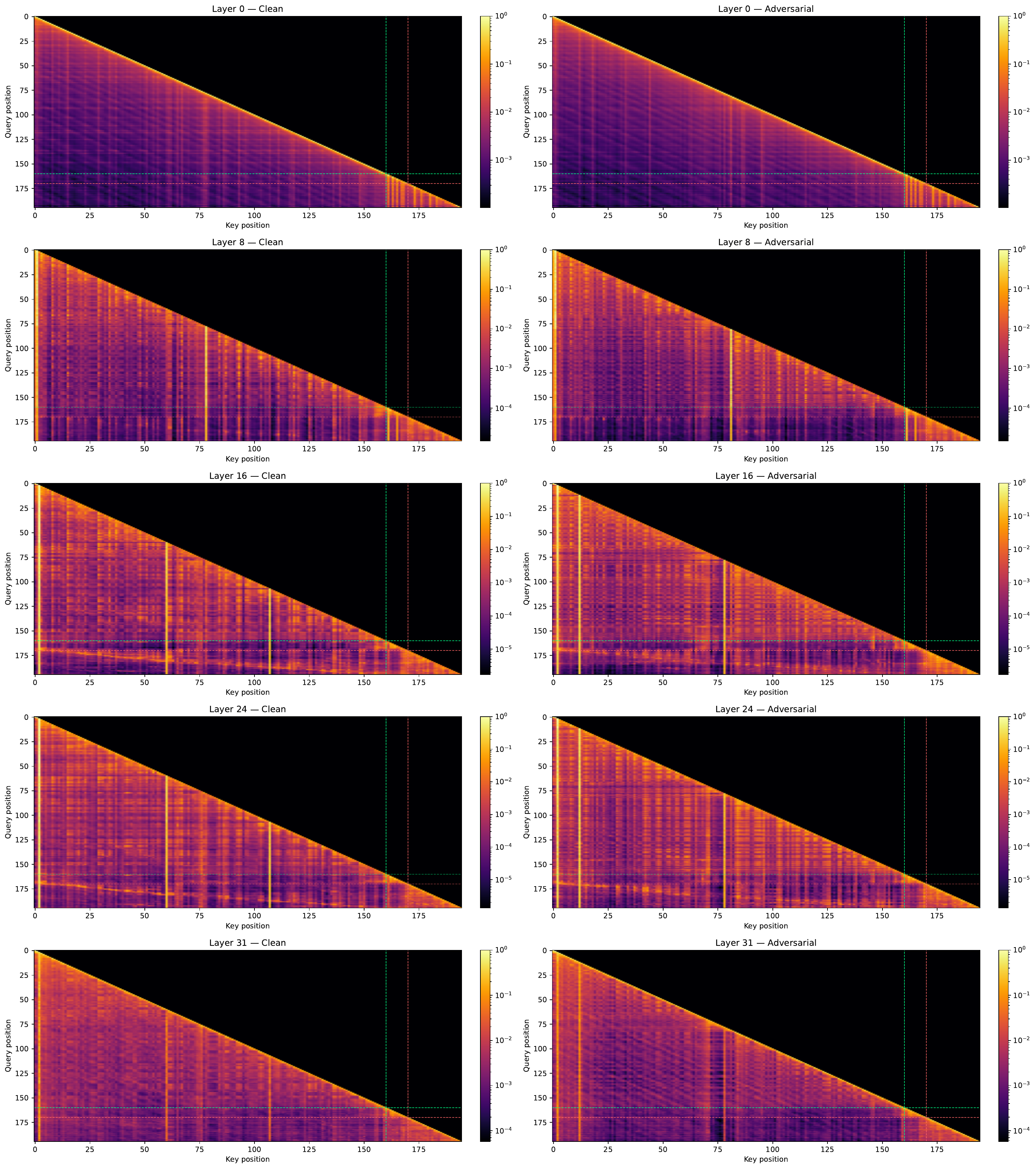}
    \caption{
    Single-sample attention maps of Qwen2-Audio under clean and adversarial inputs.
    Each row corresponds to one Transformer layer, while the left and right columns show clean and adversarial maps, respectively.
    Green dashed lines mark the audio-token boundaries, and red dashed lines mark the beginning of the target-text region.
    }
    \label{fig:single_attention_map}
\end{figure*}

Figure~\ref{fig:single_attention_map} presents a comparison between clean and adversarial attention maps. 
Brighter colors correspond to larger attention weights from the query token to the key token. To display both strong and weak attention regions, we use log-scale normalization in the visualization.

\begin{figure*}[t]
    \centering
    \includegraphics[width=\textwidth]{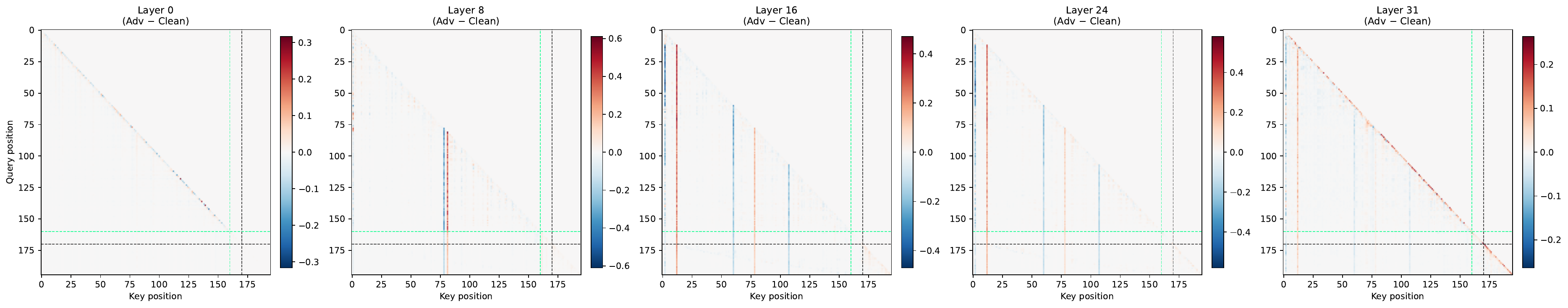}
    \caption{
    Single-sample attention difference maps between adversarial and clean inputs.
    The visualized value is $\Delta A_l = A^{\mathrm{adv}}_l - A^{\mathrm{clean}}_l$.
    Red regions indicate increased attention under adversarial input, while blue regions indicate decreased attention.
    }
    \label{fig:single_attention_diff}
\end{figure*}

To better observe the attention changes caused by the attack, we further plot the differential attention map as $\Delta A_l = A^{\mathrm{adv}}_l - A^{\mathrm{clean}}_l.$
Figure~\ref{fig:single_attention_diff} shows the differential maps across different layers.
Red indicates increased attention under the adversarial condition, while blue indicates decreased attention. 

The differential maps show that attention changes are more pronounced in the middle and upper layers, and are concentrated near the audio boundaries, the target-text start position, and several key positions strongly relied on by the model. 

Therefore, the single-sample visualization provide complementary visualizations.  Figure~\ref{fig:single_attention_map} and Figure~\ref{fig:single_attention_diff} illustrate how this change appears in specific samples, layers, and token regions. 
These observations suggest that once the semantic representations of the speech tokenizer are disrupted by our semantic attack, the cross-modal alignment process of the LALM is also affected, making it difficult for the model to recover the correct semantics from the audio context and ultimately leading to understanding degradation.

\subsection{Transcription Examples.} We provide some attack examples in Table \ref{tab:examples_1}.

\begin{table}[h]
  \centering
  \caption{Transcriptions on different ASR models}
  \label{tab:examples_1}
  \resizebox{0.9\linewidth}{!}{
  \begin{tabular}{@{}lrp{10cm}@{}}
    \toprule
    \textbf{Method} & \textbf{WER (\%)} & \textbf{Transcriptions} \\
    \midrule
    \rowcolor{gray!10}\multicolumn{3}{c}{\textbf{Whisper Large}} \\
    \midrule
    Original & -- & roaming endlessly around the park she wants to go home \\
    Clean & 0.000 & roaming endlessly around the park she wants to go home \\
    Gaussian & 10.000 & \textcolor{red}{blooming} endlessly around the park she wants to go home \\
    E2E-VGuard & 10.000 & roaming endlessly around the park \textcolor{red}{he} wants to go home \\
    AudioShield & 90.000 & \textcolor{red}{roman did not listen to the wrong that hour she} wanted to go home \\
    T-SemAttack & 250.000 & \textcolor{red}{and olympics and are on the path to a free world home have fun watching people ? have fun watching people ? have fun watching people ?} \\
    \midrule
    \rowcolor{gray!10}\multicolumn{3}{c}{\textbf{VibeVoice-ASR}} \\
    \midrule
    Original & 0.000 & i am never more aware of a room is acoustics than when i am trying to enjoy a snack i have no intention of sharing \\
    Gaussian & 12.000 & i am never more aware of a room is acoustics than when i am trying to \textcolor{red}{listen back} i have no intention of sharing \\
    E2E-VGuard & 64.000 & i am never \textcolor{red}{going to really believe the postscript and i am calling you just now} i have no intention \textcolor{red}{to} \\
    AudioShield & 28.000 & i am never more aware of a \textcolor{red}{woman} is acoustics than when i am trying to enjoy a snack i have \textcolor{red}{not looked into the tree already} \\
    T-SemAttack & 116.000 & i never \textcolor{red}{know where to leave i do not think it is going to be fun it is going to be good i hope you do not think it is too late} \\
    \midrule
    \rowcolor{gray!10}\multicolumn{3}{c}{\textbf{Qwen3-ASR}} \\
    \midrule
    Original & 0.000 & i am never more aware of a room is acoustics than when i am trying to enjoy a snack i have no intention of sharing \\
    clean & 4.000 & i am never more aware of a room is \textcolor{red}{coziest} than when i am trying to enjoy a snack i have no intention of sharing \\
    Gaussian & 4.000 & i am never more aware of a room is \textcolor{red}{coasticks} than when i am trying to enjoy a snack i have no intention of sharing \\
    E2E-VGuard & 44.000 & i am never more aware of a \textcolor{red}{wind is approach} than when \textcolor{red}{one gives me} a snack i \textcolor{red}{have made them so care} \\
    AudioShield & 28.000 & i am never more aware of \textcolor{red}{romanticostics} than when \textcolor{red}{trying to eat} a snack i have no intention of sharing \\
    T-SemAttack & 88.000 &  i \textcolor{red}{remember we are going to coast beach in lantau the green map i have used on the computer} \\
    \bottomrule
  \end{tabular}
  }
  \vspace{4pt}
\end{table}

\end{document}